\documentclass[numberedappendix,twocolumn,twocolappendix,apj]{openjournal}

\usepackage{latexsym}
\usepackage{graphicx}
\usepackage{amssymb}
\usepackage{longtable}
\usepackage{epsf}
\usepackage{amsmath}
\usepackage{graphicx}
\usepackage{hyperref}
\hypersetup{colorlinks=true,linkcolor=blue,citecolor=blue,filecolor=blue,urlcolor=blue}
\usepackage{cleveref}
\usepackage{bm}
\usepackage{lipsum}
\usepackage{enumitem}
\usepackage{txfonts}
\usepackage{gensymb}
\usepackage{booktabs}
\usepackage{multirow}
\usepackage{natbib}
\usepackage{fnpct}
\setfnpct{punct-after}

\usepackage[table]{xcolor}

\definecolor{hpurple}{HTML}{7E16DF}
\definecolor{hgreen}{HTML}{008F0F}
\definecolor{horange}{HTML}{FFA301}

\newcommand{\dd}{\rm d}
\newcommand{\eg}{\textit{e.g.}}
\newcommand{\ie}{\textit{i.e.}}
\newcommand{\panco}{\texttt{panco2}}

\newcommand{\refeq}[1]{eq.~(\ref{#1})}

\defcitealias{arnaud_universal_2010}{A10}
\newcommand{\aten}{\citetalias{arnaud_universal_2010}}

\defcitealias{bleem_cmbksz_2022}{B22}
\newcommand{\sptymap}{\citetalias{bleem_cmbksz_2022}}

\defcitealias{keruzore_exploiting_2020}{K20}
\newcommand{\nkact}{\citetalias{keruzore_exploiting_2020}}

\defcitealias{planck_collaboration_planck_2016}{P16}
\newcommand{\plckymap}{\citetalias{planck_collaboration_planck_2016}}

\begin{document}

\title{panco2: a Python library to measure intracluster medium pressure profiles from Sunyaev-Zeldovich observations}
\shorttitle{\panco: pressure profiles from tSZ maps}

\author{F. K\'eruzor\'e$^{1,2,\star}$}
\author{F. Mayet$^{2}$}
\author{E. Artis$^{2}$}
\author{J.-F. Mac\'ias-P\'erez$^{2}$}
\author{M. Mu\~noz-Echeverr\'ia$^{2}$}
\author{L. Perotto$^{2}$}
\author{F. Ruppin$^{3}$}

\affiliation{
    $^1$High Energy Physics Division, Argonne National Laboratory, 9700 South Cass Avenue, Lemont, IL 60439, USA \\
    $^2$Univ. Grenoble Alpes, CNRS, Grenoble INP, LPSC-IN2P3, 53, avenue des Martyrs, 38000 Grenoble, France \\
    $^3$Univ. Lyon, Univ. Claude Bernard Lyon 1, CNRS/IN2P3, IP2I Lyon, F-69622, Villeurbanne, France
}
\thanks{$^{\star}$E-mail:fkeruzore@anl.gov}

\shortauthors{K\'eruzor\'e et al.}


\begin{abstract}
    We present \panco, an open-source \texttt{Python} library designed to extract galaxy cluster pressure profiles from maps of the thermal Sunyaev-Zeldovich effect.
    The extraction is based on forward modeling of the total observed signal, allowing to take into account usual features of millimeter observations, such as beam smearing, data processing filtering, and point source contamination.
    \panco\ offers a large flexibility in the inputs that can be handled and in the analysis options, enabling refined analyses and studies of systematic effects.
    We detail the functionalities of the code, the algorithm used to infer pressure profile measurements, and the typical data products.
    We present examples of running sequences, and the validation on simulated inputs.
    The code is available on \href{https://github.com/fkeruzore/panco2}{github}, and comes with an extensive technical \href{https://panco2.readthedocs.io}{documentation} to complement this paper.
\end{abstract}


\maketitle

\vspace{1cm}

\twocolumngrid


\section{Introduction}
\label{sec:intro}

The millimeter domain is one of the wavelengths of choice for the detection and in-depth study of galaxy clusters.
Clusters can be observed at such frequencies through the Sunyaev-Zeldovich effect \citep[SZ,][]{sunyaev_observations_1972}, \ie\ the spectral distortion of the cosmic microwave background (CMB) due to the Compton scattering of its photons on free electrons along the line of sight.
The SZ effect is often separated in different components, depending on the origin of the energy transferred from the electrons; the main components being, by order of decreasing overall amplitude, the thermal (tSZ) and kinetic \citep[kSZ,][]{sunyaev_velocity_1980} effects \citep[see][for a recent review of the SZ effects]{mroczkowski_astrophysics_2019}.
Catalogs of clusters detected through their tSZ signal are particularly interesting for cosmological applications, as the amplitude of the tSZ effect does not suffer from cosmological dimming \citep{carlstrom_cosmology_2002}.
As a result, modern millimeter-wave sky surveys have brought us some of the largest and deepest cluster samples to date, with the catalogs built from the Atacama Cosmology Telescope \citep[ACT,][]{hilton_atacama_2021}, the South Pole Telescope \citep[SPT,][]{bleem_sptpol_2020}, and \textit{Planck} \citep{planck_collaboration_planck_2016-2} surveys.

The amplitude of the tSZ distortion is directly proportional to the electron pressure in the gaseous intracluster medium (ICM) integrated along the line of sight.
This link between tSZ signal and ICM pressure motivates studies of the pressure distribution in the ICM -- in its simplest form, as a spherically symmetric pressure profile.
For example, matched-filtering cluster detection algorithms may require a prior assumption on the overall shape of the ICM pressure profile \citep[\eg][]{melin_comparison_2012}, in which case a poor knowledge of this property of clusters may be an obstacle to the construction of a cluster sample. 
Similarly, the interpretation of the angular power spectrum of the tSZ signal over the whole sky, which can be used to constrain cosmology, strongly relies on an assumption of the pressure profile of clusters, and the recovering of cosmological parameters can be severely affected by its poor knowledge \citep{ruppin_impact_2019}.
The mean pressure profile of galaxy clusters has been investigated using different cluster samples over the last decade.
Early works conducted on local, X-ray selected samples, such as \citet[][hereafter \aten]{arnaud_universal_2010}, converged towards a ``universal'' pressure profile, undergoing self-similar redshift and mass evolution \citep[see also \eg][]{plagge_sunyaev-zeldovich_2010,battaglia_cluster_2012-1, planck_collaboration_planck_2013}.
In these studies, the main determining factor for the shape of the pressure profile of a cluster was its dynamical state, with relaxed clusters exhibiting a steeper pressure profile in their core.
Nevertheless, the pressure profile of clusters is expected to deviate from self-similarity due to various baryonic processes, such as feedback by active galactic nuclei jets and supernovae explosions \citep[\eg][]{nagai_effects_2007}.
The impact of these processes is still largely unknown, as they are expected to have a larger importance in the shallower gravitational potentials of low-mass halos, which are harder to detect and study observationally.
To better understand these impacts, large efforts are made to produce hydrodynamical simulations with a rich description of baryonic physics \citep[\eg][]{cui_three_2018-1,villaescusa-navarro_camels_2021,pakmor_millenniumtng_2022}.
 These simulations in turn need to calibrate their subgrid models on observation-based studies, making the measurement of pressure profiles from observations a key element of studies of cluster physics and cosmology \citep[\eg][]{dolag_sz_2016,ruppin_impact_2019-1,wadekar_sz_2022}.

The first step in any evaluation of the mean pressure profile of a sample is the extraction of individual profiles from observed data.
Such measurements can be performed from X-ray cluster observations, using a deprojection of the ICM electron density and temperature \citep[see \eg][for reviews]{bohringer_x-ray_2010, bohringer_x-ray_2013}.
Because of cosmological dimming, the X-ray surface brightness measure towards a cluster -- at fixed density -- strongly decreases with redshift, with $S_{\rm X} \propto (1+z)^{-4}$.
As a consequence, very high integration times may be required to get X-ray observations deep enough to infer quality measurements of pressure profiles of high redshift systems.
Alternatively, one may use tSZ observations of clusters, which do not suffer from redshift dimming, enabling the detection of more distant objects.
In particular, high angular resolution millimeter observations of clusters with large aperture telescopes have successfully been used to measure cluster pressure profiles, and are today one of the preferred sources of data for studies of the mean pressure profile of clusters \citep[\eg][]{pointecouteau_pact_2021,sayers_evolution_2022,young_mean_nodate}.

In this paper, we present \panco, a \texttt{Python} library developped to perform pressure profile extraction from tSZ observations.
The algorithm is based on forward modeling of the tSZ signal and MCMC sampling, and allows users to account for different features of millimeter-wave observations that may manifest as systematic biases or uncertainties in recovered pressure profiles.
An earlier version of \panco\ was described in \citet{keruzore_panco2_2022}, which offered less flexibility in the analysis, as the only data that could be analyzed was maps from the NIKA2 camera 150~GHz channel \citep{adam_nika2_2018, perotto_calibration_2020}.
This software represents an evolution of the software developed within the NIKA and NIKA2 collaborations, and has already been used for different studies based on NIKA2 data \citep[\eg][]{artis_psz2_2022,munoz-echeverria_multi-probe_2022,munoz-echeverria_lpsz-clash_2022}.
Here, we present a generalization of the code, that makes it able to perform pressure profile extractions from arbitrary tSZ datasets.

This paper is structured as follows.
We present the algorithm implemented in \panco\ to estimate pressure profiles from tSZ maps in \S\ref{sec:algo}.
In \S\ref{sec:simu}, we present a set of realistic simulated maps, mimicking observations by different millimeter-wave instruments of mock clusters at different masses and redshifts, that we use to validate the accuracy of the pressure profiles recovered by \panco.
In \S\ref{sec:extra_simu}, we extend the validation dataset to incorporate more complex data features, such as point source contamination and anisotropic filtering.
We discuss the caveats of \panco\ and some possibilities for improvements in \S\ref{sec:discussion}, and conclude in \S\ref{sec:conclusion}.

Throughout this paper, even though \panco\ can use different cosmological models, we assume a flat $\Lambda {\rm CDM}$ Universe, with $\Omega_{\rm m}=0.3, \;\Omega_\Lambda=0.7, \; h=0.7$.
This cosmology is mainly used to infer angular diameter distances to the cluster being studied from its redshift, in order to map sky distances to physical ones, as well as to generate mock pressure profiles using the cosmology-dependent universal profile of \aten.
Quantities with a $500$ subscript refer to the properties of a cluster within its characteristic radius $R_{500}$, corresponding to the radius of a sphere around the center of the cluster in which the mean matter density is $500$ times the critical density of the Universe at the cluster's redshift.

\section{Algorithm} \label{sec:algo}

\begin{figure*}[htp]
    \centering
    \includegraphics[width=.95\linewidth]{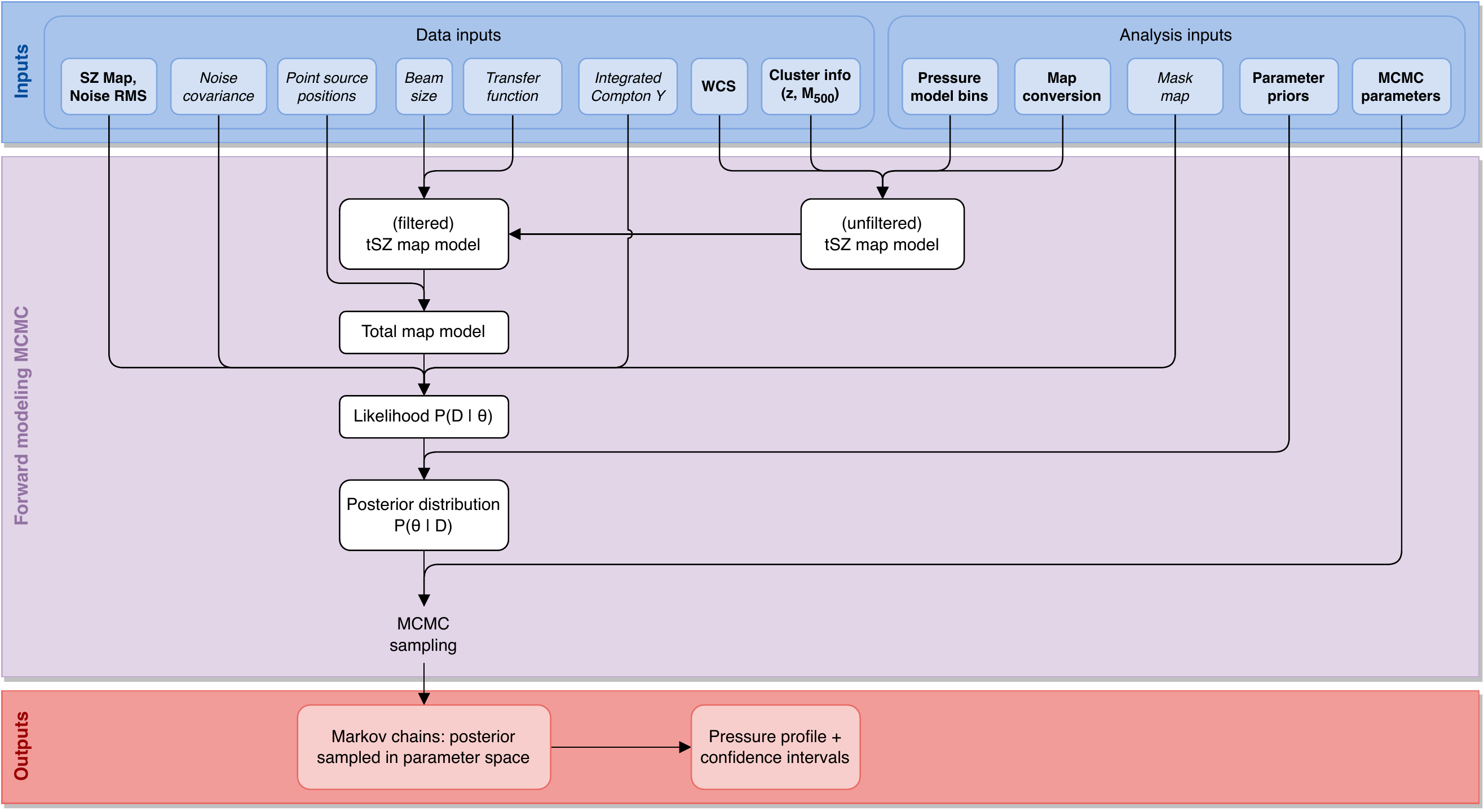}
    \caption{
        Schematic workflow of the \panco\ algorithm, from its inputs (blue), to the forward modeling and MCMC sampling (purple), and results (red).
        Required and optional inputs are denoted with boldface and italic fonts, respectively.
    }
    \label{fig:workflow}
\end{figure*}

The goal of \panco\ is to infer an estimate of a pressure profile and of its confidence intervals from the tSZ map of a cluster.
The overall workflow implemented in \panco\ to perform this measurement is presented in Figure~\ref{fig:workflow}.
It is based on the forward modeling of the SZ map and on Monte-Carlo Markov Chain (MCMC) sampling of the probability distribution for the pressure profile parameters given the input data.
In this section, we detail each step of the analysis, as well as the inputs to be given to \panco\ and the results it produces.

\subsection{Inputs} \label{sec:algo:inputs}

The main input of \panco\ is a mapping of a patch of the sky containing tSZ signal.
The code reads in a file in the FITS format \citep{wells_fits_1981} for its input data, that must include the following elements:

\begin{itemize}[leftmargin=*]
    \item An extension in which the data is the SZ map to be fitted, and the header includes the World Coordinate System (WCS) used to create the map, in order to correctly map sky coordinates to pixels using the flat-sky approximation;
    \item An extension in which the data represent an estimate of the expected root mean squared (RMS) error for each pixel of the data map.
\end{itemize}

Such a file constitutes the minimum data input for \panco\ to be able to fit a pressure profile.
Using these inputs, the user may choose to only use a square portion of the map, by specifying the sky coordinates of its center and its side.

Additional inputs can be provided to account for various data features.
\paragraph{Beam smearing} the user may provide the width of a Gaussian function to account for point spread function (PSF, hereafter referred to as ``beam'') filtering (see \S\ref{sec:algo:fwdmod});
\paragraph{Transfer function} Fourier filtering due to data processing and/or scanning strategy can be accounted for in the analysis (see \S\ref{sec:algo:fwdmod});
\paragraph{Point source contamination} the position on the sky of point sources, as well as an \textit{a priori} probability distribution for their fluxes, can be used to account for the contamination and marginalize over its amplitude (see \S\ref{sec:algo:fwdmod});
\paragraph{Correlated noise} the covariance matrix between the noise of pixels in the map can be provided (see \S\ref{sec:algo:likelihood}).
\panco\ also offers routines to compute the covariance from various data inputs, such as the power spectrum of the residual noise;
\paragraph{Integrated SZ signal} an external measurement of the integrated Compton parameter, that may be used as a constraint on large-scale tSZ signal, can be given to \panco\ (see \S\ref{sec:algo:likelihood});
\paragraph{Mask} arbitrary parts of the maps may be masked in the analysis, \eg\ to avoid contamination by point sources or to separately extract profiles for different sectors in the map \citep[as in \eg][]{ruppin_unveiling_2020}.

\subsection{Pressure profile model} \label{sec:algo:press}

The electron pressure distribution in the ICM is modeled in \panco\ as a radial pressure profile, implying spherical symmetry of the ICM.
The most widely used parametrization for ICM pressure profiles, called the generalized Navarro-Frenk-White model \citep[gNFW,][]{zhao_analytical_1996, nagai_effects_2007}, is known to have several shortcomings.
In particular, the very self-constrained shape of the functional form of the profile and the important correlations in the parameter space make model fitting complex, and the recovered parameter values hard to interpret \citep[see \eg][]{nagai_effects_2007, battaglia_cluster_2012-1, sayers_evolution_2022}.

To try to circumvent these issues, \panco\ uses a more flexible parametrization of the pressure profile, in which the pressure distribution is modeled as a power law evolution in concentric spherical shells.
In this modeling, the model parameters are the values $P_i$ of the pressure profile at predefined radii from the cluster center $R_i$, with a power law interpolation between the radii:
\footnote{Several authors also refer to this modeling as a non-parametric model.}
\begin{equation}
    \label{eq:algo:pressure_profile}
    P(r \in [R_i, R_{i+1})) = P_i \left(r / R_i\right)^{-\alpha_i},
\end{equation}
where $\alpha_i = - \log(P_{i+1} / P_i) / \log(R_{i+1} / R_i)$. \\
Outside the radial bins, the pressure profile is extrapolated using the power-law evolution of the first (last) bin.
In addition to being more flexible and having generally lesser correlations in the parameter space, this parametrization can be integrated along a line of sight analytically through an Abel transform, by following \eg\ \citet{basu_non-parametric_2010, romero_multi-instrument_2018}.

The main downside to this pressure profile modeling lies in the need to specify the radii $R_i$ used in \refeq{eq:algo:pressure_profile} \textit{a priori} when performing a fit.
There is no obvious, fail-safe way to define this radial binning.
A user may choose a binning motivated by the data coverage (\eg\ with radii that, projected on the sky plane, are separated by the size of the beam of the instrument used to build the map); but looking at the same data, another user may be motivated by sample studies over several clusters, and wish to define a binning in units of a characteristic radius of the cluster (\eg\ $R_{500}$).
We note that for the same input data, results might differ between several choices of radial binning; this model dependence of the results will be addressed in \S\ref{sec:discussion}.

\subsection{Forward modeling: from pressure profile to SZ map} \label{sec:algo:fwdmod}

\begin{figure}
    \centering
    \includegraphics[height=0.933\textheight]{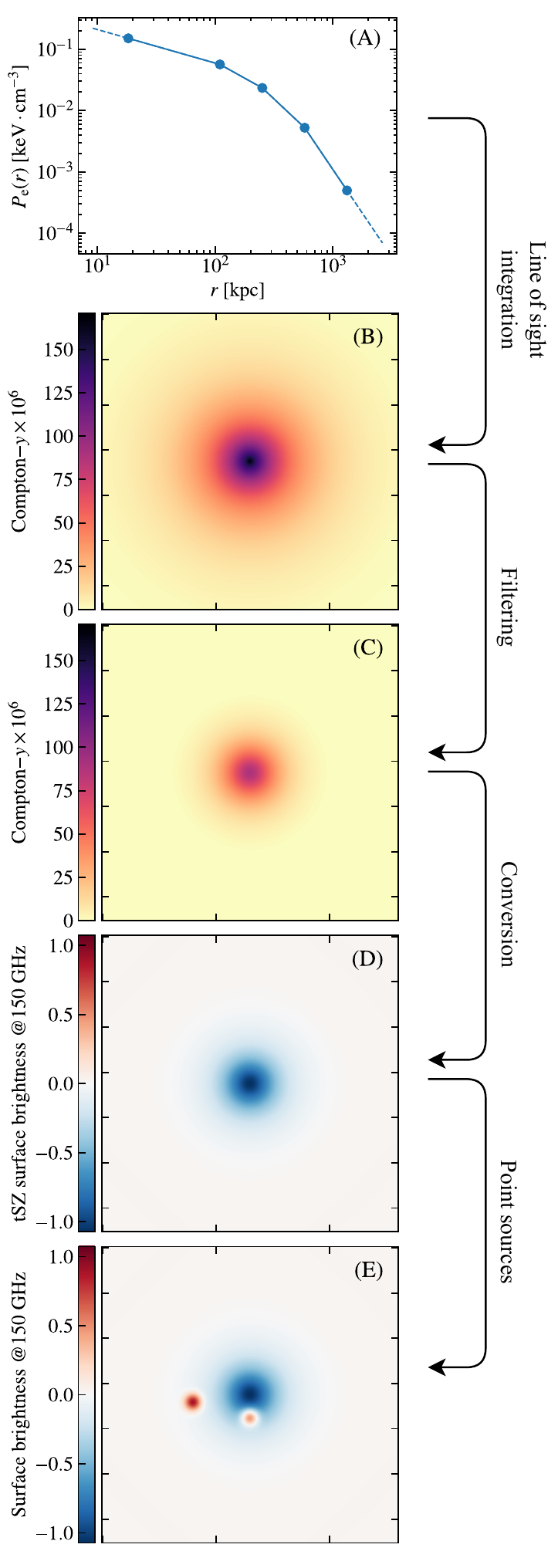}
    \caption{
        Illustration of the forward modeling procedure.
        The pressure profile (A) is integrated along the line of sight to create a Compton$-y$ map (B), which is filtered (C) and converted (D) to be realistically comparable to the observed data.
        If point source contamination information is passed, point source models can be added to the map (E).
    }
    \label{fig:fwmod}
\end{figure}

The approach used by \panco\ to fit pressure profiles on SZ maps is forward modeling.
In that framework, a pressure profile model -- determined by \refeq{eq:algo:pressure_profile} -- is used to generate a map that can be compared to data.
This approach has been vastly used in the estimation of pressure profiles from tSZ maps, especially in the context of resolved follow-up of galaxy clusters with \eg\ NIKA \citep[\eg][]{adam_first_2014,romero_multi-instrument_2018,ruppin_non-parametric_2017}, NIKA2 \citep[\eg][]{ruppin_first_2018,keruzore_exploiting_2020,munoz-echeverria_multi-probe_2022}, MUSTANG(2) \citep[\eg][]{romero_galaxy_2017,romero_pressure_2020}, Bolocam \citep[\eg][]{sayers_evolution_2022}, or ALMA \citep[\eg][]{di_mascolo_joint_2019}.
This section details the different steps used in that process, which is illustrated in Figure~\ref{fig:fwmod}.

\paragraph{Line of sight integration}

The amplitude of the tSZ effect in a direction $\theta$ on the sky is named the Compton parameter $y$, and is proportional to the integral of the electron pressure along the line of sight (LoS):
\begin{equation}
    \label{eq:algo_ysz}
    y(\theta) = \frac{\sigma_\textsc{t}}{m_{\rm e}c^2} \int_{\rm LoS(\theta)} P_{\rm e} \; \dd l,
\end{equation}
where $\sigma_\textsc{t}$ is the Thompson cross-section, and $m_{\rm e} c^2$ is the electron resting energy. \\
In \panco, we perform this integration analytically, by following the derivation presented in Appendix A of \citet{romero_multi-instrument_2018}, using the spherically symmetric case. 
This allows us, for any given pressure profile, to create a Compton parameter map in the same coordinate system as the data, \ie\ an estimate of the value of $y$ for each pixel in the map.

\paragraph{Conversion and zero level}

Depending on the data product available, and on the convention used in the raw data processing software employed to create these data products, tSZ maps can be expressed in a variety of units (\eg\ surface brightness, CMB temperature fluctuation).
These units can usually be converted to Compton$-y$ through a scalar conversion coefficient, $C_{\rm conv}$, which can depend on many different quantities that may be difficult to estimate, or even fluctuate during observations, such as instrumental bandpasses, weather conditions, instrumental calibration, or even temperature of the ICM through relativistic corrections to the tSZ effect \citep{mroczkowski_astrophysics_2019}.
As a result, the conversion coefficient is affected by an uncertainty.
In \panco, this coefficient is treated as a parameter in the model, for which a prior distribution needs to be specified (see \S\ref{sec:algo:mcmc}), allowing one to propagate the uncertainty on the conversion of the map to the pressure profile estimate.
This means that a vector in the parameter space will contain a value of a conversion coefficient, by which \panco\ multiplies the model $y-$map to get a map in the same units as the input data.
In case the input data is already in units of Compton$-y$, this coefficient may still be used to propagate multiplicative calibration uncertainty by centering the prior distribution on~1.
In addition, in order to enable taking into account possible large-scale residual noise, a zero-level offset $Z$ can be added to the map, and marginalized over.

\paragraph{Filtering}

The tSZ maps constructed by any instrument are affected by different types of filtering.
The instrumental PSF acts as a filter that smooths the data by suppressing signal at small scales.
In its forward modeling approach, \panco\ is able to take into account this filtering by convolving the model $y-$map with a Gaussian filter, the width of which can be specified by the user.

In addition, during the data reduction process used to create tSZ maps from raw data, filtering can occur, often suppressing signal at large angular scales.
This filtering is usually accounted for through a transfer function, quantifying the signal filtering in the Fourier space, and evaluated during the data processing.
\panco\ can account for this effect by filtering the map with a transfer function in Fourier space, to be provided by the user.
Two different types of transfer functions can be provided:
\begin{itemize}[leftmargin=*]
    \item a 1D transfer function: assuming that the filtering is isotropic, \panco\ can convolve model maps with a 1D kernel specified by the user as multipole moments $\ell$ and their associated filtering ${\rm TF}(\ell)$;
    \item a 2D transfer function: if the filtering on the sky cannot be assumed to be isotropic, the user may specify 2D multipoles $(\ell_x, \ell_y)$ and their corresponding filter ${\rm TF}(\ell_x, \ell_y)$.
\end{itemize}
A thorough discussion of the impact of choosing a 1D or 2D transfer function -- in the case of NIKA2 data, in which the filtering is mildly anisotropic -- can be found in \citet{munoz-echeverria_multi-probe_2022}.

\paragraph{Point source contamination}

The resulting \panco\ model map contains the tSZ model signal, in the units of and mapped as the input data, and affected by the same signal filtering.
It is therefore comparable to the tSZ signal in the input map.
But millimeter-wave maps of cluster regions may also include astrophysical signal from other sources.
In particular, signal from millimeter-emitting galaxies can often be found in such maps.
These galaxies can be foreground, background, or cluster members, and their emission can be thermal (in the case of dusty star-forming galaxies), or synchrotron (for radio-loud AGN).
In any case, this signal manifests as a contamination for tSZ science, and must be accounted for in data analysis, lest the recovered pressure profile be biased.

In its forward modeling approach, \panco\ uses the methodology described in \eg\ \citet[and references therein]{keruzore_exploiting_2020}, in which point source models may be added to the tSZ model map in order to model the total signal in the map.
The spatial model of each source is given by the instrumental PSF, and their fluxes are treated as model parameters, with priors specified by the user (see \S\ref{sec:algo:mcmc}).
If the flux of a source is well known from external data, this prior will be tight, and serve as a propagation of its uncertainty to the recovered pressure profile.
Otherwise, if a source is known to be present but little is known about its flux, large priors can be used, in which case \panco\ will constrain the sum of the tSZ signal and the point source flux\footnote{\panco\ is able to constrain the sum of the two because of the different spatial distribution of the tSZ and point source fluxes.}.
To account for point source contamination in the analysis, the user must therefore provide \panco\ with a position (in sky coordinates) and a probability distribution for the flux (in input map units) for each of the sources considered.

\subsection{Likelihood} \label{sec:algo:likelihood}

The process described in \S\ref{sec:algo:fwdmod} allows \panco\ to compute a model map that is comparable to the input data from any position in the parameter space.
To summarize, a vector in this parameter space, $\vartheta$, has the following components:
\begin{itemize}[leftmargin=*]
    \setlength\itemsep{0pt}
    \item $P_0, \dots P_n$: the value of the pressure profile at each predefined radius $R_0, \dots R_n$ ;
    \item $C_{\rm conv}$: the conversion coefficient from Compton$-y$ to input map units;
    \item $Z$: a zero-level for the map;
    \item If provided, $F_0, \dots F_m$: the fluxes of point sources in the map, in input map units.
\end{itemize}

The comparison between the model map and the input data is performed through a Gaussian likelihood function:
\begin{align}
    \nonumber -2 \log \mathcal{L}(\vartheta) & \equiv -2 \log p(D | \vartheta) + {\rm cst.} \\
        & = \left[D - M(\vartheta)\right]^{\rm T} {\rm C}^{-1} \left[D - M(\vartheta)\right],
    \label{eq:algo:likelihood}
\end{align}
where $D$ is the input map, $M(\vartheta)$ is the model map computed from the position in parameter space $\vartheta$, and ${\rm C}$ is the noise covariance matrix. \\
If a mask is provided in the analysis, the masked pixels are not considered in the likelihood computation, \ie\ $D$ and $M$ are the masked data and model map, and ${\rm C}$ only includes the covariance between unmasked pixels.

If the noise in the map can be considered white, the noise values in the map pixels are uncorrelated, and ${\rm C}$ becomes diagonal.
In that case, in order to reduce the computation time needed, the likelihood is rewritten as:
\begin{equation}
    \label{eq:algo:likelihood_white}
    -2 \log \mathcal{L}(\vartheta) = \sum_i \left( \frac{D_i - M_i(\vartheta)}{\Sigma_i} \right)^2,
\end{equation}
where the sum runs over all unmasked pixels $i$ in the map, and $\Sigma$ is the uncertainty associated to each pixel,  that can be given by the noise RMS map.

In some cases, especially with high angular resolution follow-ups of clusters, additional information may be known from SZ surveys, such as the integrated tSZ signal $Y_{<R}$ within a radius $R$ of the cluster.
This knowledge can serve as an additional constraint on the model, by computing the value of the integrated tSZ signal from the model as the spherical integration of the pressure profile within the radius $R$.
This value can then be compared to that known from independent measurement, providing an additional data point that can be added to the log-likelihood of eqs.~(\ref{eq:algo:likelihood} or \ref{eq:algo:likelihood_white}):
\begin{equation}
    \label{eq:algo:likelihood_ysz}
    \left(\frac{Y_{<R}^{\rm meas.} - 4\pi\frac{\sigma_\textsc{t}}{m_{\rm e}c^2}\int_{0}^{R} P_{\rm e}(r) \, r^2 \dd r}{\Delta Y_{<R}^{\rm meas.}} \right)^2,
\end{equation}
where $Y_{<R}^{\rm meas.}$ and $\Delta Y_{<R}^{\rm meas.}$ are the measured integrated SZ signal within $R$ and its uncertainty. \\
This additional constraint can be especially useful when dealing with observations agressively filtering large scale signal, in which case it might effectively provide constraining power on the pressure profile at large radii.
This approach is routinely used for NIKA2 tSZ follow-ups of SZ-detected clusters \citep[\eg][]{ruppin_first_2018,keruzore_exploiting_2020,munoz-echeverria_multi-probe_2022}.

\subsection{Posterior distribution and MCMC sampling} \label{sec:algo:mcmc}

\panco\ performs the extraction of a pressure profile from tSZ data using Bayesian MCMC.
In that framework, a prior probability distribution for the parameters, $p(\vartheta)$, must be multiplied to the likelihood function of \refeq{eq:algo:likelihood} to obtain a posterior distribution for the parameters given the data: $p(\vartheta | D) \propto p(D | \vartheta) \, p(\vartheta)$.
In \panco, the prior distributions for the different model parameters are considered uncorrelated, meaning the prior distribution is the product of the individual priors on parameters:
\begin{equation}
    \label{}
    p(\vartheta) = \prod_i p(\vartheta_i),
\end{equation}
where the product runs over all individual parameters $i$. \\
The prior on each parameter is to be specified by the user, using the large variety of distributions available in the \texttt{scipy.stats}\footnote{\url{https://docs.scipy.org/doc/scipy/reference/stats.html}} module \citep{virtanen_scipy_2020}.

The resulting posterior distribution, $p(\vartheta | D)$, is then sampled using MCMC.
More specifically, we use the affine-invariant ensemble sampling implementation of the \texttt{emcee} library \citep{foreman-mackey_emcee_2019}.
Convergence of the chains is monitored at regular intervals based on the autocorrelation length of the chains, using the following algorithm.
Every $n_{\rm check}$ steps (\ie\ accepted positions in the parameter space), the integrated autocorrelation length $\tau_j$ of each chain $j$ is computed, as well as the average autocorrelation over all chains, $\tau = \left< \tau_j \right>$.
Convergence is accepted if both of the two following criteria are met:
\begin{enumerate}[leftmargin=*]
    \item The current length of the chains is longer than $n_{\rm auto} \tau$;
    \item The mean autocorrelation has changed by less than $\Delta\tau_{\rm max}$ in the last two evaluations.
\end{enumerate}
The values of $n_{\rm auto}$ and $\Delta\tau_{\rm max}$ are parameters of \panco, that need to be specified by the user.
Likewise, the user needs to provide a maximum chain length at which the sampling should stop if convergence was never reached.

\subsection{Chains cleaning and exploitation} \label{sec:algo:outputs}

Once MCMC convergence has been reached, \panco\ stores the full sampling of the posterior distribution, \ie\ all of the accepted positions in the parameter space and their associated log-likelihood and log-posterior values.
The raw chains can then be loaded and cleaned as follows:
\begin{enumerate}[leftmargin=*]
    \item Remove the first $n_{\rm burn}$ samples as a burn-in length;
    \item Thin the chains by discarding $(n_{\rm discard} - 1) / n_{\rm discard}$ samples, \ie\ only keeping one sample every $n_{\rm discard}$ steps;
    \item Discard the chains that are poorly mixed, \ie\ systematically outside of the $[q_{\rm extr}, 1 - q_{\rm extr}]$ quantiles of the sampled posterior.
\end{enumerate}
Again, the values of $n_{\rm burn}$, $n_{\rm discard}$ and $q_{\rm extr}$ are parameters of the analysis, that must be user-provided.

The cleaned chains can then be expressed in the pressure profile space.
To do so, \refeq{eq:algo:pressure_profile} is used to compute a profile for each position in the parameter space, over a radial range specified by the user.
These profiles may then be saved for future analyses.

The Markov chains, in the parameter space and in the pressure profile space, constitute the main data product of \panco.
Several figures can be produced for a visual representation of the results.
They are all presented in the online technical documentation\footnote{\url{https://panco2.readthedocs.io/en/latest/notebooks/example_C2_NIKA2.html}}.
Here we give a brief description of these figures.

\paragraph{Posterior distribution}
Several functions are available to produce figures showing properties of the Markov chains and the posterior distribution they sample, using the \texttt{ChainConsumer} library \citep{hinton_chainconsumer_2016}:
\begin{itemize}[leftmargin=*]
    \item Walks plot, \ie\ the evolution of the positions in the parameter space with the number of steps for each parameter;
    \item Corner plot, \ie\ a figure showing the marginalized posterior for all individual parameters and sets of two parameters.
        The prior distribution can be overplotted for comparison purposes;
    \item Plot of the correlation and covariance matrices of the different parameters.
\end{itemize}

\paragraph{Data -- Model -- Residuals}
Two figures can be produced to illustrate the quality of the fit.
They both consist in comparing the input data, the best-fitting model, and the residuals (\ie\ the difference between input data and best-fitting model).
These can be plotted in 2D, by showing the three maps side-by-side (see Figure~\ref{fig:valid:dmr_2d}), or in 1D, by showing the radial profiles of each of these maps in the same graph (see the left panels of Figure~\ref{fig:valid:profiles}).
Goodness of fit can be judged visually in both representations by searching for significant structure in the residuals.

\paragraph{Pressure profile}
In addition, a plot of the median of the pressure profiles computed from each posterior sample, as well as confidence intervals chosen as their 16th and 84th percentiles, can be produced (see the right panels of Figure~\ref{fig:valid:profiles}).

\section{Validation on simulations} \label{sec:simu}

In order to ensure that \panco\ is able to recover accurate pressure profile measurements, we test it on simulated inputs.
In this section, we detail this validation process, from the creation of the synthetic dataset to the results produced by \panco.
For reproducibility purposes, the datasets created and used for this analysis are made public with the software.

\subsection{Sample selection}

The goal of the validation is to ensure that \panco\ is able to recover accurate pressure profiles from different types of data.
To that end, we seek to create realistic synthetic cluster maps from three instruments: the \textit{Planck} satellite, the South Pole Telescope (SPT), and the NIKA2 camera at the IRAM 30-m telescope.
The choice of these three instruments is motivated by their vastly different angular resolutions: the Compton$-y$ maps built from \textit{Planck} and SPT data have angular resolutions (expressed as the full width at half maximum, or FWHM) of $10'$ and $1.25'$, respectively \citep{planck_collaboration_planck_2016, bleem_cmbksz_2022}, and the beam of the NIKA2 camera 150 GHz band -- used for tSZ mapping -- has an FWHM of $18''$ \citep{perotto_calibration_2020}.

\begin{figure}[tp]
    \centering
    \includegraphics[width=\linewidth]{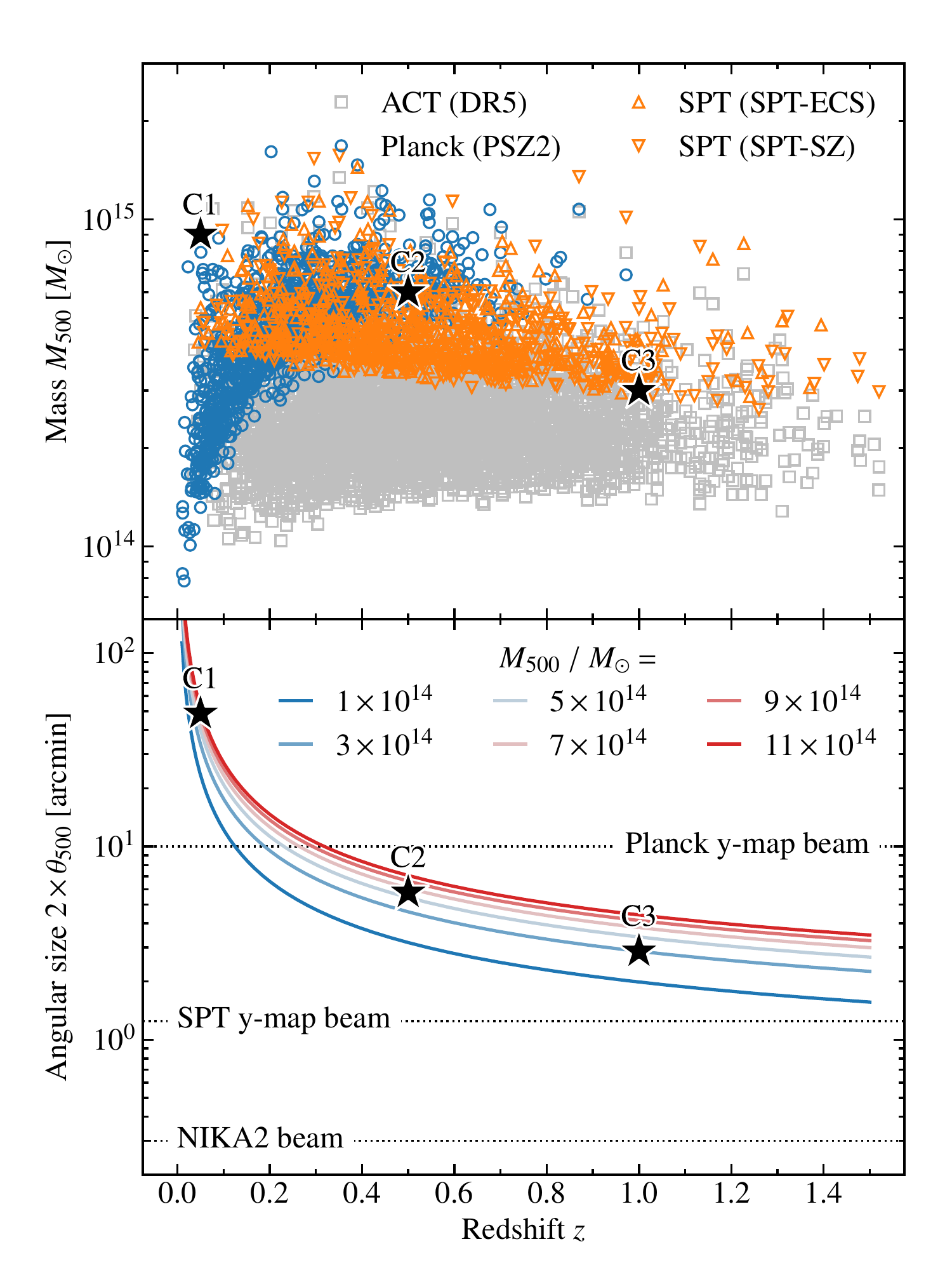}
    \caption{
        Validation cluster sample (black stars) in the mass-redshift plane (\textit{top panel}) and in the angular size-redshift plane (\textit{bottom panel}).
        For illustration, the top panel includes clusters detected in recent tSZ surveys: \textit{Planck} \citep{planck_collaboration_planck_2016-2}, ACT \citep{hilton_atacama_2021}, and SPT \citep{bleem_sptpol_2020,bleem_galaxy_2015}.
        Note that the redshift axis is truncated to $z<1.6$, therefore not showing all clusters in these samples.
        Colored lines in the bottom panel show the evolution of the angular $2\theta_{500}$ with redshift for clusters of different masses.
        The angular resolutions of the \textit{Planck} and SPT $y-$maps, as well as that of the NIKA2 camera at 150~GHz, are represented as dotted horizontal lines.
    }
    \label{fig:valid:sample}
\end{figure}

We choose to create SZ maps for three clusters, labeled (C1, C2, C3), covering different regions of the mass-redshift plane:
\begin{alignat}{2}
    \nonumber {\rm C1}:\; & z=0.05,\ & M_{500} = 9 \times 10^{14} \ M_\odot ;\\
    \nonumber {\rm C2}:\; & z=0.5,\  & M_{500} = 6 \times 10^{14} \ M_\odot ;\\
              {\rm C3}:\; & z=1,\    & M_{500} = 3 \times 10^{14} \ M_\odot,
    \label{eq:valid:clusters}
\end{alignat}
These mock clusters are shown as black stars in Figure~\ref{fig:valid:sample}.
The top panel shows their positions in the mass-redshift plane, indicating that C1, C2 and C3 are realistic detections for the \textit{Planck}, SPT, and ACT tSZ surveys, respectively.
The bottom panel of Figure~\ref{fig:valid:sample} places the clusters in the angular diameter-redshift plane, showing that C1 can be resolved in \textit{Planck}, SPT, and NIKA2 tSZ maps, while C2 and C3 are too small to be resolved by \textit{Planck}.

\subsection{Data generation} \label{sec:simu:mkdata}

\begin{table*}[htp]
    \centering
    \begin{tabular}{l c c c c c}
        \toprule
        Instrument & Map size & Pixel size & FWHM & Filtering & White noise \\
        (Clusters) & $\theta_{\rm map}$ & $\theta_{\rm pix}$ & $\theta_{\rm FWHM}$ & & level \\
        \midrule
        \textit{Planck} & \multirow{2}{*}{$5\degree$} & \multirow{2}{*}{$2'$} & \multirow{2}{*}{$10'$} & \multirow{2}{*}{--} & Homogeneous \\
        (C1) & & & & & RMS=$4.12 \times 10^{-6} \; [y]$ \\
        \midrule
        SPT & \multirow{2}{*}{$30'$} & \multirow{2}{*}{$15''$} & \multirow{2}{*}{$1.25'$} & -- & Homogeneous, \\
        (C1, C2) & & & & & RMS=$9.78 \times 10^{-6} \; [y]$ \\
        \midrule
        NIKA2 & \multirow{2}{*}{$6.5'$} & \multirow{2}{*}{$3''$} & \multirow{2}{*}{$18''$} & NIKA2-like & Isotropic \\
        (C2, C3) & & & & transfer function & NIKA2-like RMS \\
        \bottomrule
    \end{tabular}
    \caption{\normalfont
        Properties of simulated maps emulating cutouts of the \textit{Planck} \citep{planck_collaboration_planck_2016} and SPT \citep{bleem_cmbksz_2022} $y-$maps, and NIKA2 cluster observations \citep{keruzore_exploiting_2020}.
    }
    \label{tab:simu:map_props}
\end{table*}

We create mock maps for the three clusters by forward modeling their tSZ signal.
We assume that each cluster has a pressure profile that follows the universal profile of \aten, scaled with its mass and redshift.
This pressure distribution is integrated along the line of sight to obtain a Compton$-y$ map.
This map is then projected on a flat-sky grid and convolved with a Gaussian filter to account for instrumental filtering, and added to a random noise realization.

As discussed previously, and illustrated in the bottom panel of Figure~\ref{fig:valid:sample}, the angular size of the chosen clusters varies from a few arcminutes to several tens of arcminutes, thus their resolved mapping is not achievable by the all the instruments.
We therefore create different sets of maps for the three clusters.
For C1, the most extended cluster, we create mock \textit{Planck} and SPT maps.
For C2, our intermediate case, we create SPT and NIKA2 mock maps.
Finally, for C3, the smallest cluster in our sample, we only create mock NIKA2 data.
The characteristics of the maps are designed to mimic known cluster observations by the three instruments, and summarized in table~\ref{tab:simu:map_props}.
The generated maps are shown on the left panels of Figure~\ref{fig:valid:dmr_2d}.

\paragraph{Planck-like data}  
Our mock \textit{Planck} dataset is designed to emulate the \textit{Planck} \texttt{MILCA} \citep{hurier_milca_2013} Compton$-y$ map of \citet[hereafter \plckymap]{planck_collaboration_planck_2016}.
The original data products for this map are in \texttt{healpix} format, which \panco\ cannot process, as it relies on the flat-sky approximation.
We therefore create maps of $(5\degree \times 5\degree)$ patches of the sky using a gnomonic projection, with a pixel size of $2'$, and a Gaussian PSF of ${\rm FWHM} = 10'$.
The noise in these maps is white and has a homogeneous distribution, with an RMS taken from the left panel of Figure~13 in \plckymap.
We do not consider any filtering of the signal aside from the beam convolution, as the scales filtered out by the data processing are very large compared to the size of the patch considered (see \plckymap).

\paragraph{SPT-like data}  
Our SPT-like maps mimic the publicly available $y-$map released by the SPT collaboration \citep[][hereafter \sptymap]{bleem_cmbksz_2022}, specifically the ``minimum variance'' map.
We use the same projection as their flat-sky maps, \ie\ a Sanson-Flamsteed projection with $15''$ pixels, on a $(30' \times 30')$ patch of the sky.
We use a Gaussian beam with ${\rm FWHM} = 1.25'$.
We inject white noise realizations, the amplitude of which is evaluated by computing the RMS of the map in a $(5\degree \times 5\degree)$ patch of the sky, taking care of masking sources.
The SPT $y-$maps use information from the \textit{Planck} $y-$map in the Fourier space to compensate for large-scale filtering, making it negligible ($\leqslant 1.2\%$, from \sptymap). 
Therefore, we choose to consider no large-scale filtering in the creation of these maps.

\paragraph{NIKA2-like data}  
Our NIKA2-like data imitates the NIKA2 150~GHz sky maps obtained by the NIKA2 SZ Large Program \citep{mayet_cluster_2020, perotto_nika2_2022}.
In particular, we use the data products publicly released in \citet[hereafter \nkact]{keruzore_exploiting_2020}.
We create a gnomonic projection of a $(6.5' \times 6.5')$ patch of the sky with $3''$ pixels.
The angular resolution of the map is Gaussian with ${\rm FWHM} = 18''$, and we also take into account filtering of angular scales due to data processing via the transfer function of \nkact.
The noise in these NIKA2 maps is considered white and isotropic, but not homogeneous, as the scanning strategy used for observations of the NIKA2 SZ Large Program creates a variation in the noise level of the maps depending on the distance from the pointing center.
To accurately take into account this effect, we use the noise RMS map of \nkact, which we multiply by a scalar value to account for different exposure times.
Data produced by the NIKA2 collaboration is in NIKA2 150 GHz surface brightness units, usually presented in mJy/beam.
The conversion coefficient of $y-$to$-$mJy/beam can be evaluated by integrating the tSZ spectral distortion in the NIKA2 150 GHz effective bandpass (accounting for atmospheric opacity at the time of the observations), and is usually of the order of $-12$ mJy/beam/$y$, therefore we choose this value for our map generation.

\subsection{Pressure profile fitting} \label{sec:simu:fit}

Each of the five generated maps is then used to extract a pressure profile using \panco, following the procedure detailed in \S\ref{sec:algo}.
We assume that we know the characteristics of the map prior to fitting -- \ie\ the width of the beams used, the transfer function, and the noise RMS given to \panco\ in input are the same ones that have been used to generate the maps.

Two ingredients in our model then remain to be specified: the radial binning -- \ie\ the value of the radii $R_i$ in eq.~(\ref{eq:algo:pressure_profile}) -- and the priors on the parameters.
For the radial binning, we choose one that is identically determined by the angular coverage of each map.
The first bin is defined as the projected radius corresponding to the size of a map pixel, $R_0 = \mathcal{D}_{\rm A}(z) \tan^{-1} \theta_{\rm pix}$, where $\mathcal{D}_{\rm A}(z)$ is the angular diameter distance to the cluster redshift $z$.
Four bins $\left\{R_1 \dots R_4 \right\}$ are then added, log-spaced between the projected sizes of the beam FWHM, $\mathcal{D}_{\rm A}(z) \tan^{-1} \theta_{\rm FWHM}$, and of the half map size, $\mathcal{D}_{\rm A}(z) \tan^{-1} \theta_{\rm map} / 2$.
We stress that the choice of a first bin at such a small radius is due to the extrapolation needed to define the radially-binned profile.
As discussed in \S\ref{sec:algo:press} and \refeq{eq:algo:pressure_profile}, the pressure for all radii $r \in (0, R_0]$ is given by the extrapolation of the power law profile between radii $R_0$ and $R_1$.
As a consequence, placing a bin at small radii allows us to anchor the constraints at all radii smaller than $R_1$.
Nonetheless, with this approach, $R_0$ corresponds to a projected radius that is smaller than the resolution of the map.
Reconstructed pressure profiles at such small radii should therefore be treated with caution.

The priors on each parameter is defined as follows.
For the pressure parameters $P_i$, corresponding to the values of the pressure profile at radii $R_i$, we compute the value of the universal pressure profile from \aten\ for the cluster's mass and redshift.
The prior on $P_i$ is then set as a log-uniform distribution around this value:
\begin{equation}
    \label{eq:simu:prior_Pi}
    p(P_i) = \log\mathcal{U}(10^{-2}, 10^2) \times P_{\rm A10}(R_i).
\end{equation}
The prior on the conversion coefficient is set as a Gaussian distribution.
For \textit{Planck}- and SPT-like maps, the data is directly in units of Compton$-y$, therefore this distribution is centered around 1.
For NIKA2 data, as discussed in \S\ref{sec:simu:mkdata}, the central value of the prior is set at $-12$ mJy/beam/$y$.
For all data, the spread of the distribution is taken as $5\%$ of its central value.

The MCMC sampling is run as presented in \S\ref{sec:algo:mcmc}, with $30$ walkers, and setting $n_{\rm auto} = 50$ and $\Delta\tau_{\rm max} = 5\%$.
For our tests, we used a 2021 MacBook Pro with an M1 Pro chip and a 10-core CPU, with the MCMC using 5 threads.
With this setup and these criteria, each fit took 5 to 10 minutes to converge.
The raw Markov chains are then saved, and cleaned for the exploitation as described in \S\ref{sec:algo:outputs}, with $n_{\rm burn} = 500$, $n_{\rm discard} = 20$, and $q_{\rm extr} = 20\%$.

\subsection{Results} \label{sec:simu:results}

\begin{figure*}[p]
    \centering
    \rotatebox[x=10pt, y=58pt]{90}{C1, \textit{Planck}}
    \includegraphics[height=4cm, trim={1.5cm 0cm 1.5cm 1.75cm}, clip]{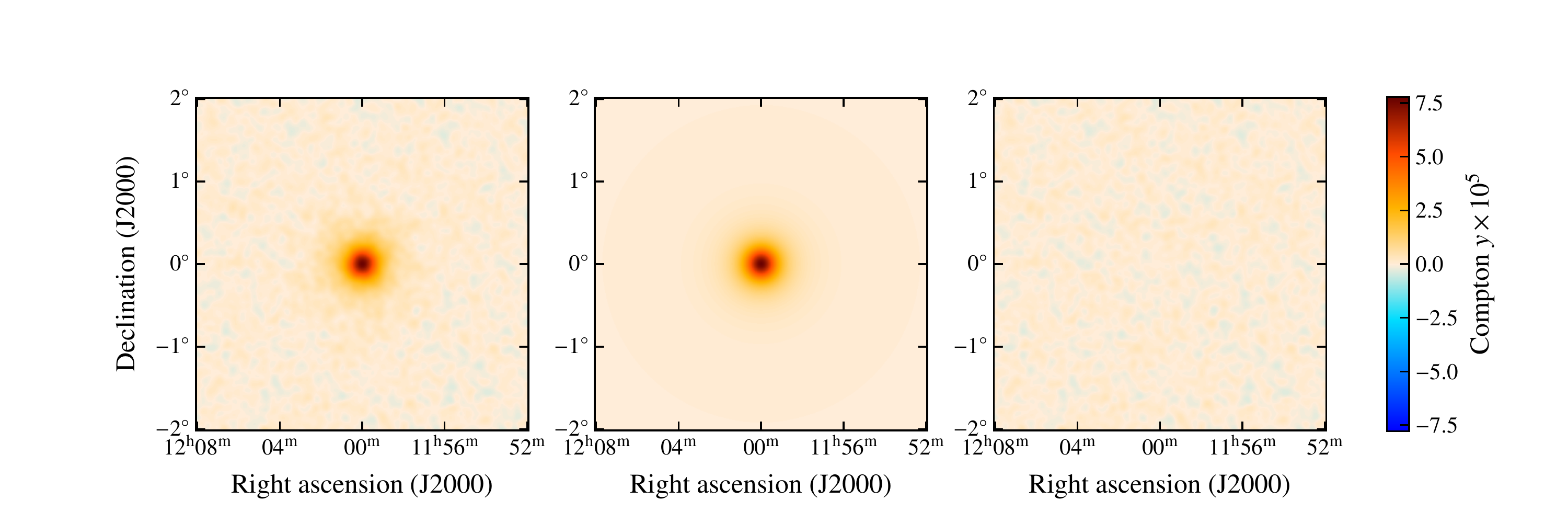} \\
    \rotatebox[x=10pt, y=65pt]{90}{C1, SPT}
    \includegraphics[height=4cm, trim={1.5cm 0cm 1.5cm 1.75cm}, clip]{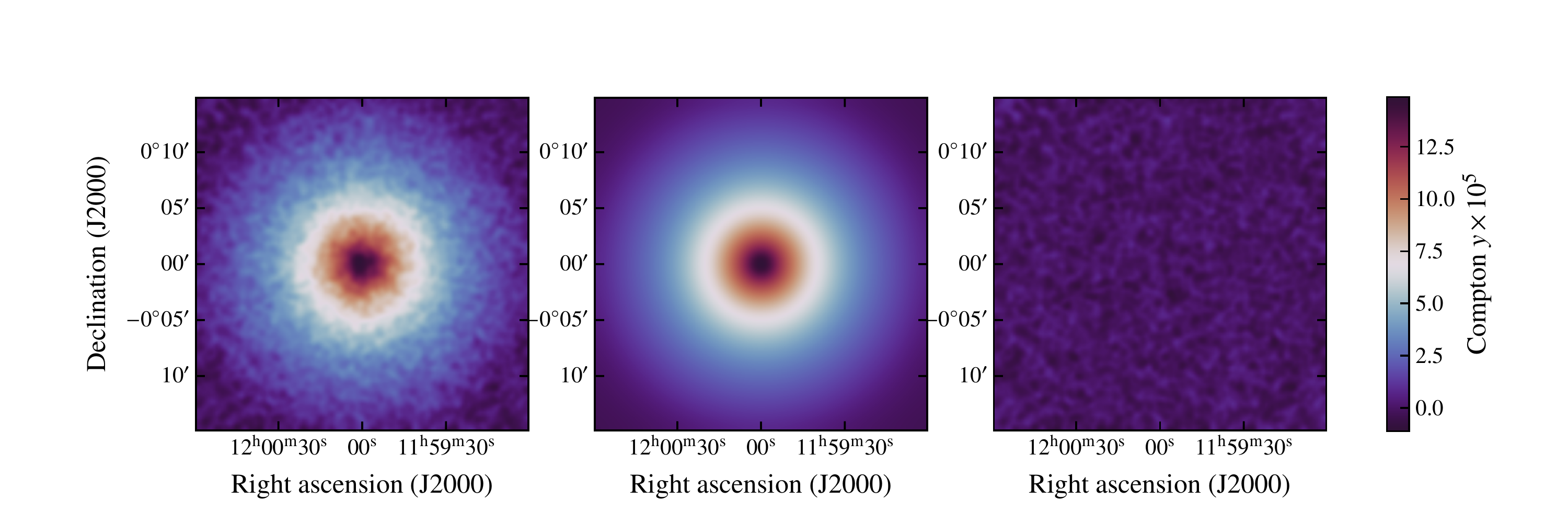} \\
    \rotatebox[x=10pt, y=65pt]{90}{C2, SPT}
    \includegraphics[height=4cm, trim={1.5cm 0cm 1.5cm 1.75cm}, clip]{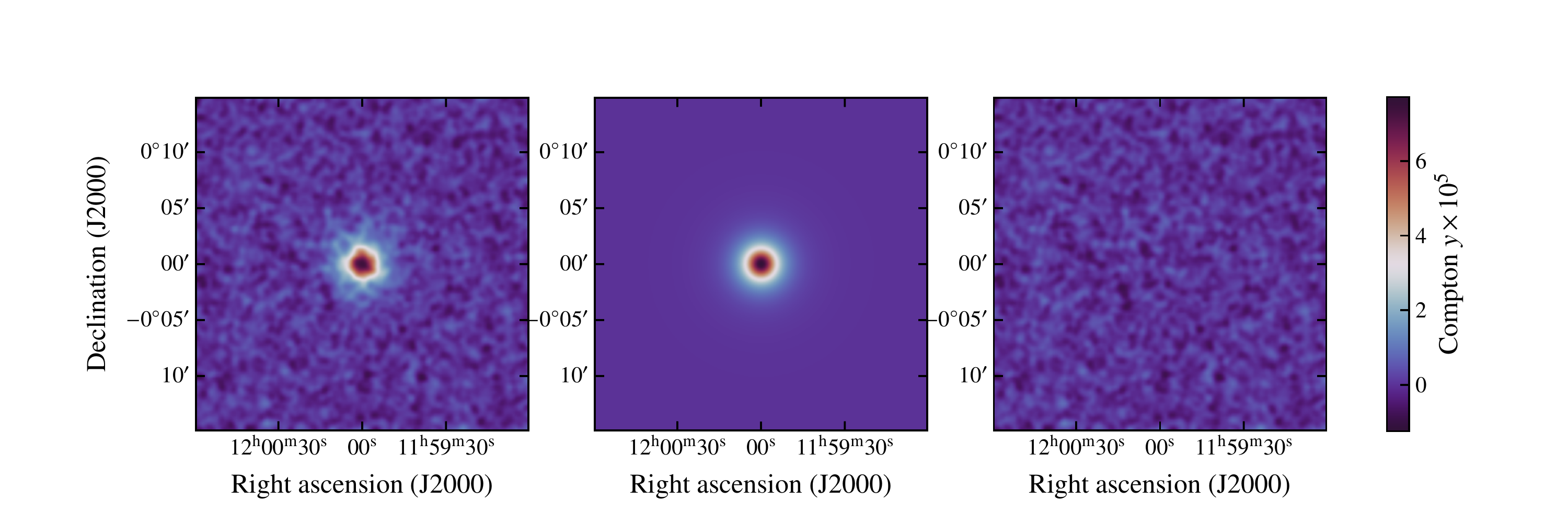} \\
    \rotatebox[x=10pt, y=60pt]{90}{C2, NIKA2}
    \includegraphics[height=4cm, trim={1.5cm 0cm 1.5cm 1.75cm}, clip]{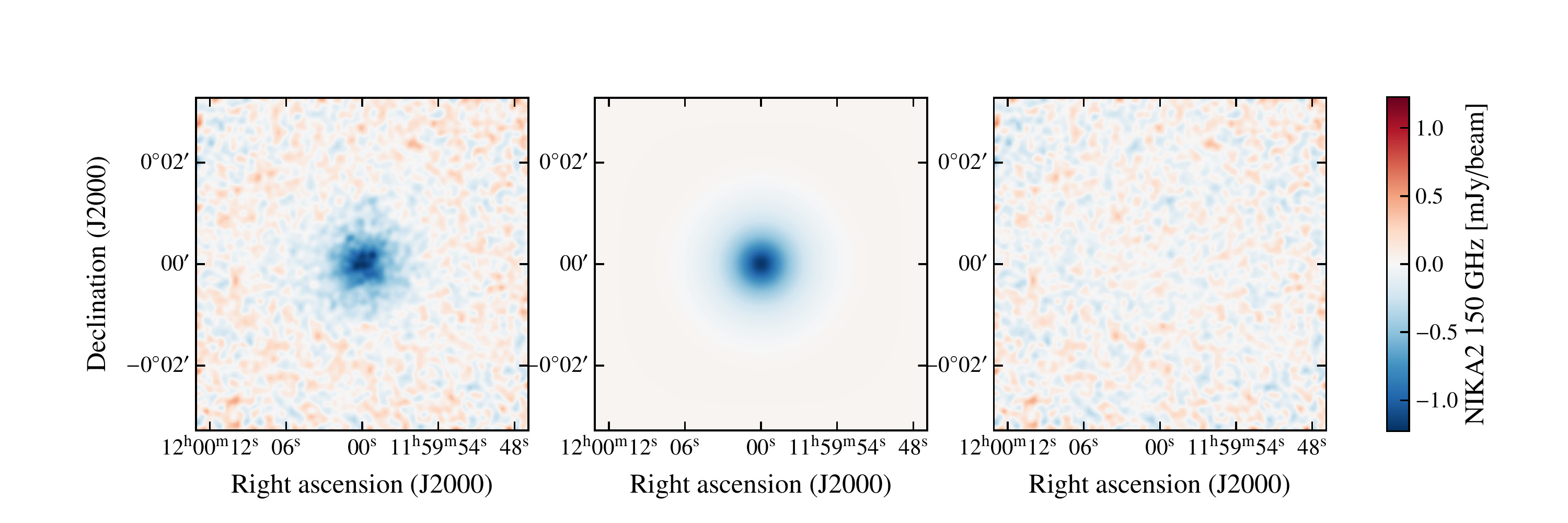} \\
    \rotatebox[x=10pt, y=60pt]{90}{C3, NIKA2}
    \includegraphics[height=4cm, trim={1.5cm 0cm 1.5cm 1.75cm}, clip]{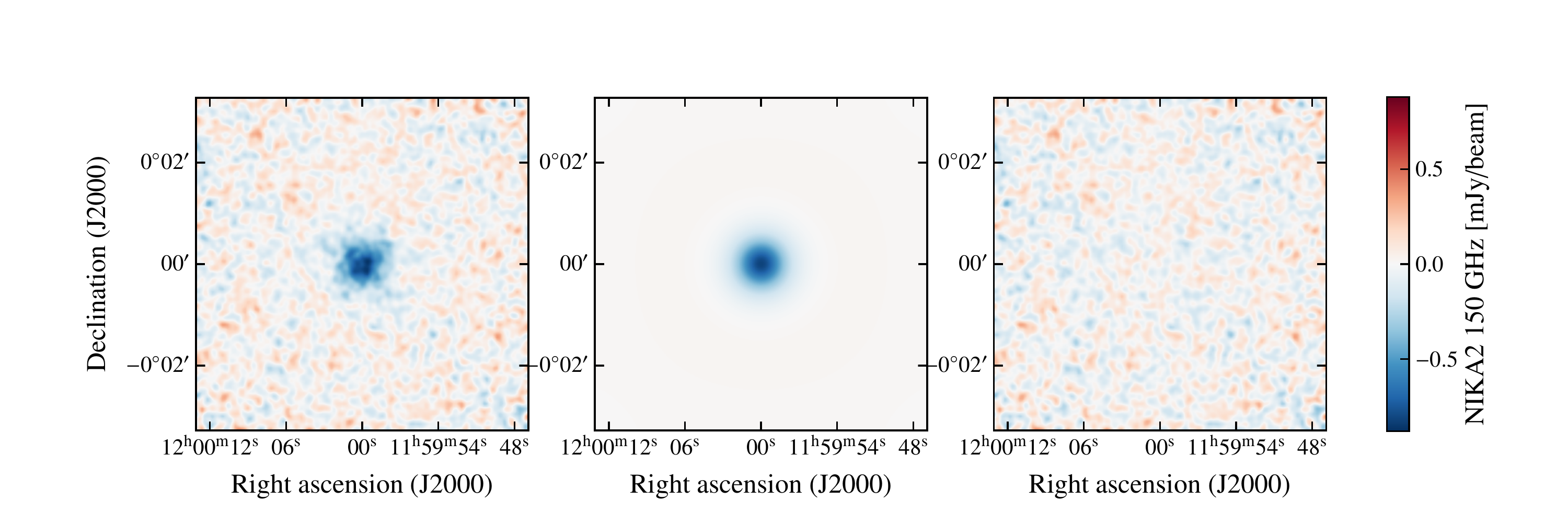}
    \caption{
        Data (\textit{left}), model (\textit{center}), and residuals (\ie\ difference between data and model, \textit{right}) for the five validation fits:
        from top to bottom, C1 with \textit{Planck}, C1 with SPT, C2 with SPT, C2 with NIKA2, C3 with NIKA2.
        Maps are smoothed by a $1$ pixel Gaussian for display purposes.
    }
    \label{fig:valid:dmr_2d}
\end{figure*}

\begin{figure*}[p]
    \centering
    \rotatebox[x=10pt, y=55pt]{90}{C1, \textit{Planck}}
    \includegraphics[height=4cm, trim={0 0 0 1cm}, clip]{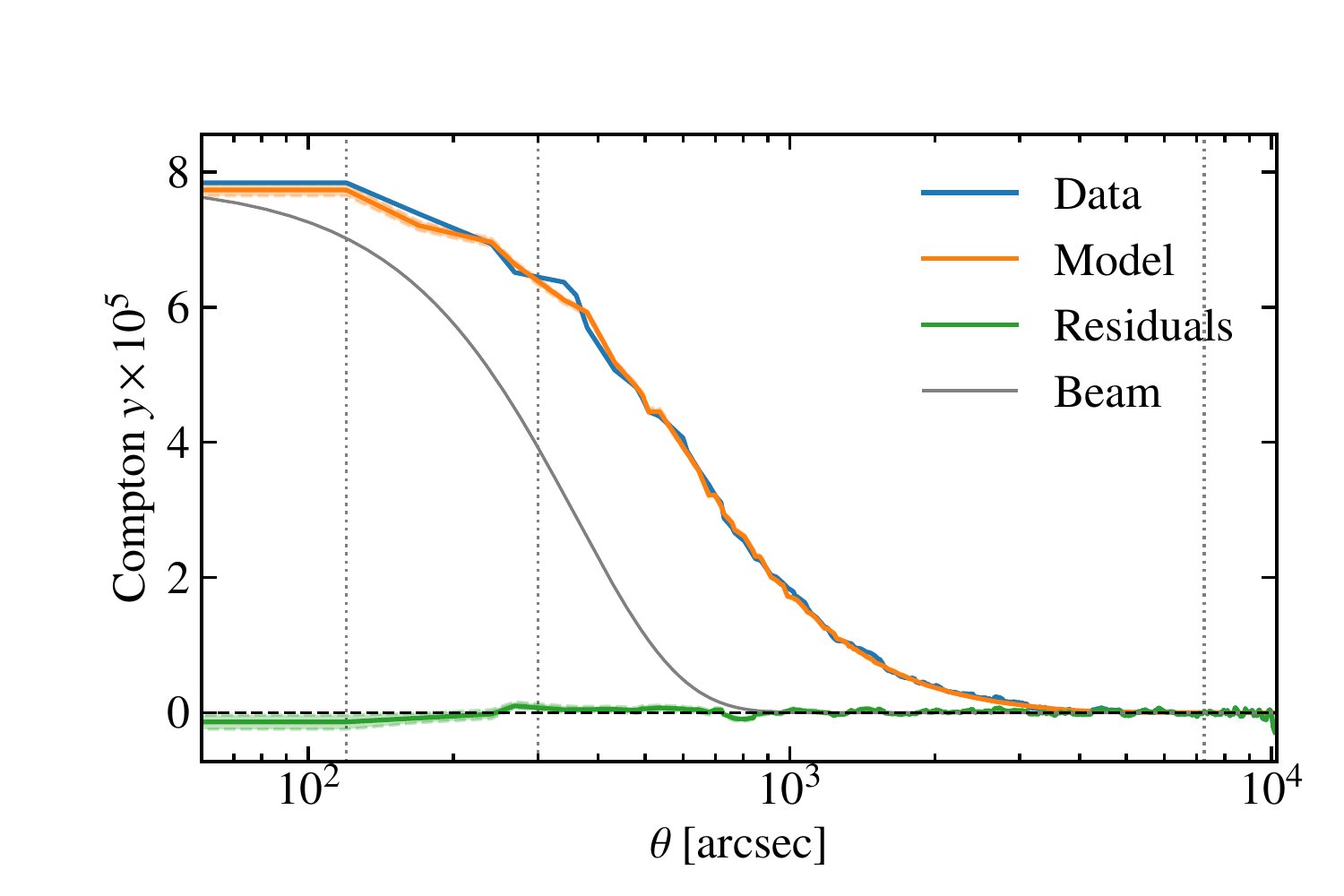}
    \includegraphics[height=4cm, trim={0 0 0 1cm}, clip]{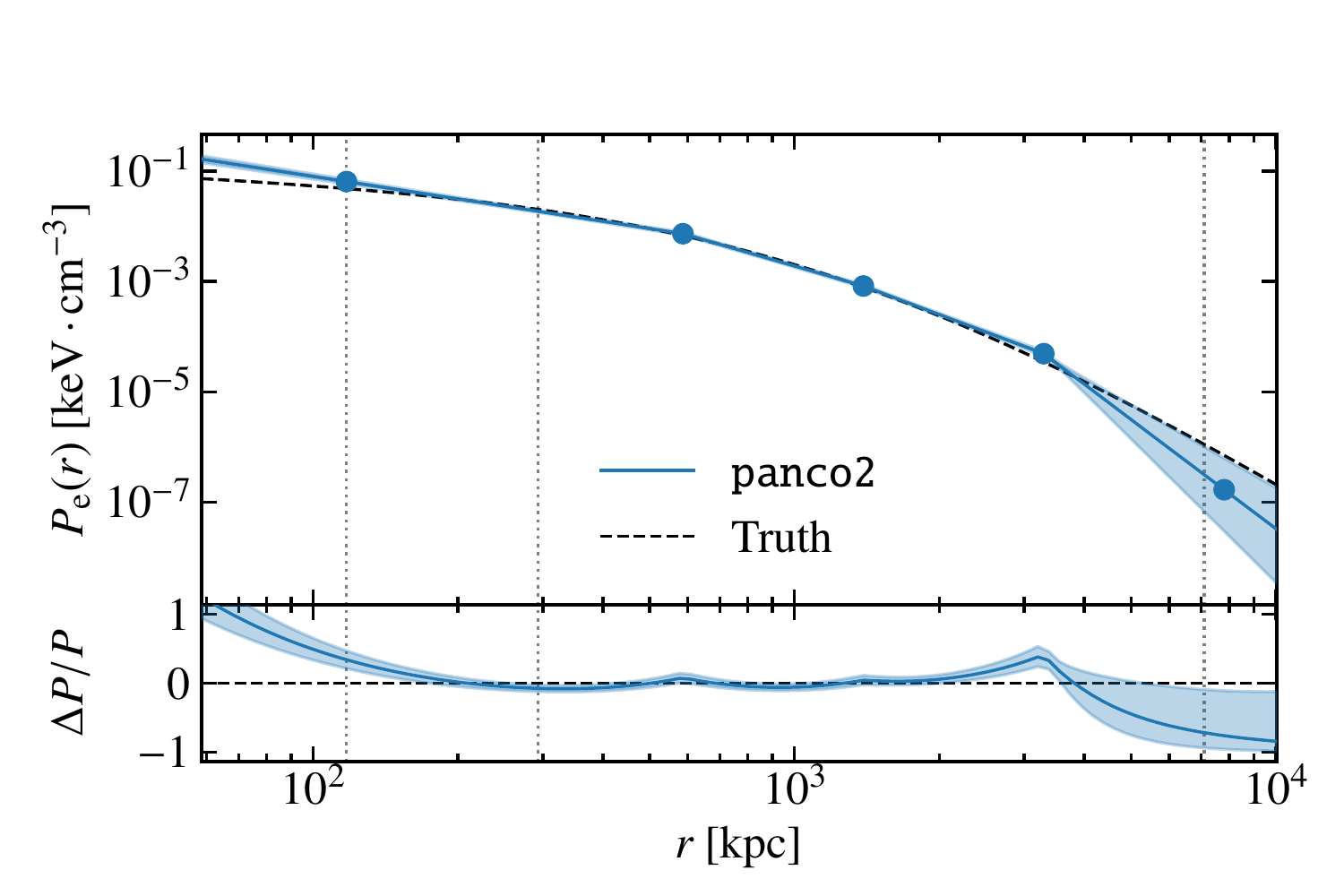} \\
    \rotatebox[x=10pt, y=60pt]{90}{C1, SPT}
    \includegraphics[height=4cm, trim={0 0 0 1cm}, clip]{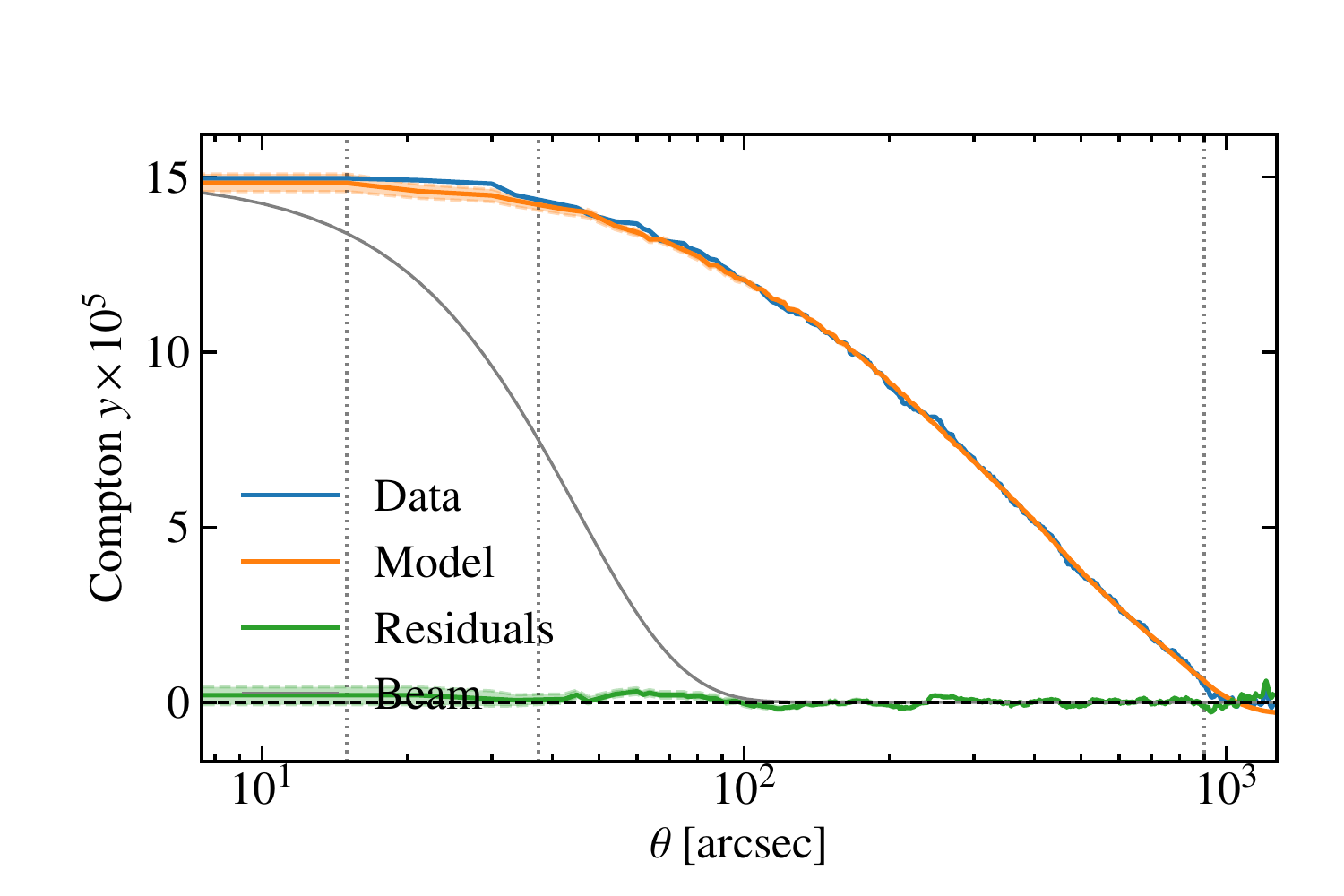}
    \includegraphics[height=4cm, trim={0 0 0 1cm}, clip]{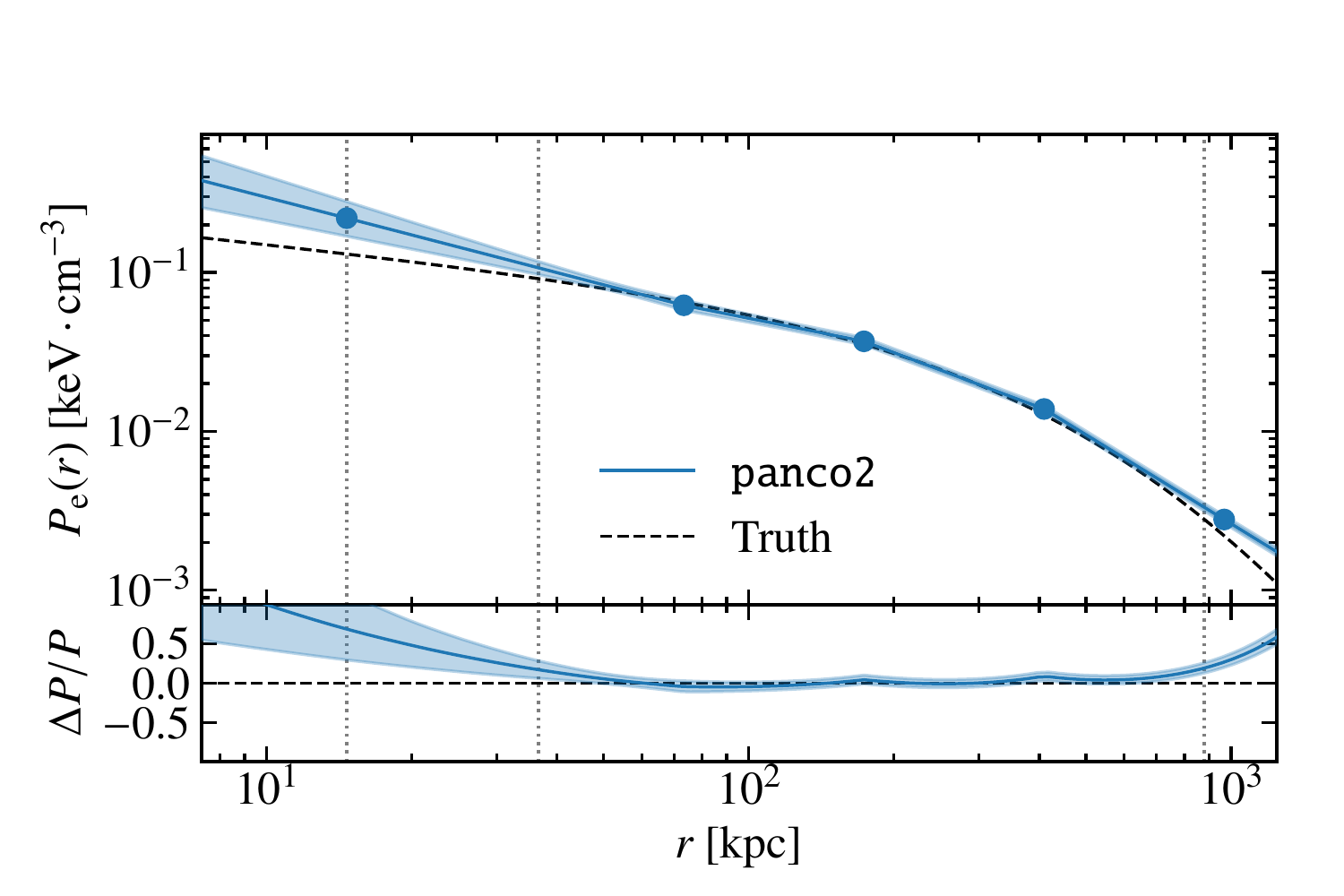} \\
    \rotatebox[x=10pt, y=60pt]{90}{C2, SPT}
    \includegraphics[height=4cm, trim={0 0 0 1cm}, clip]{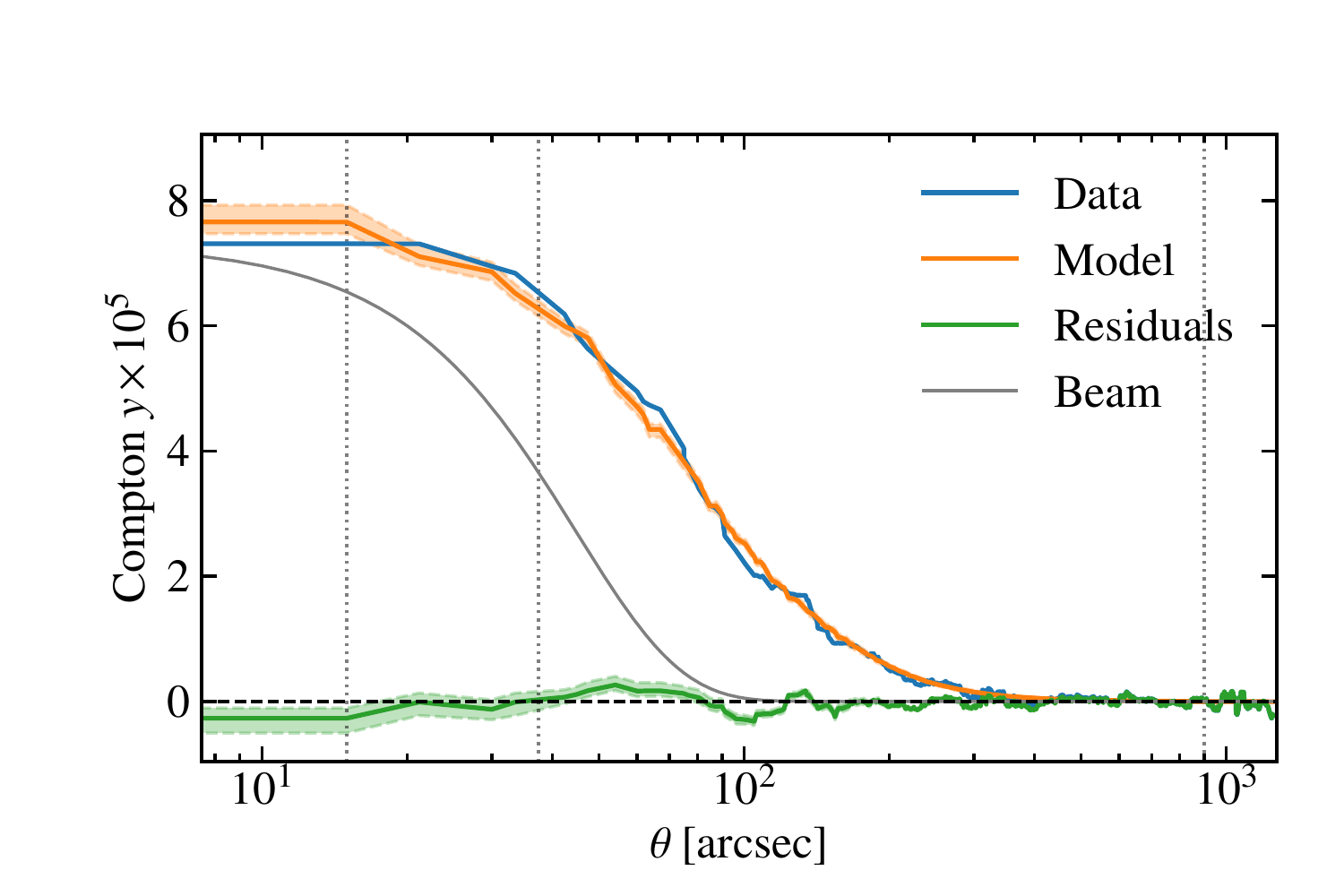}
    \includegraphics[height=4cm, trim={0 0 0 1cm}, clip]{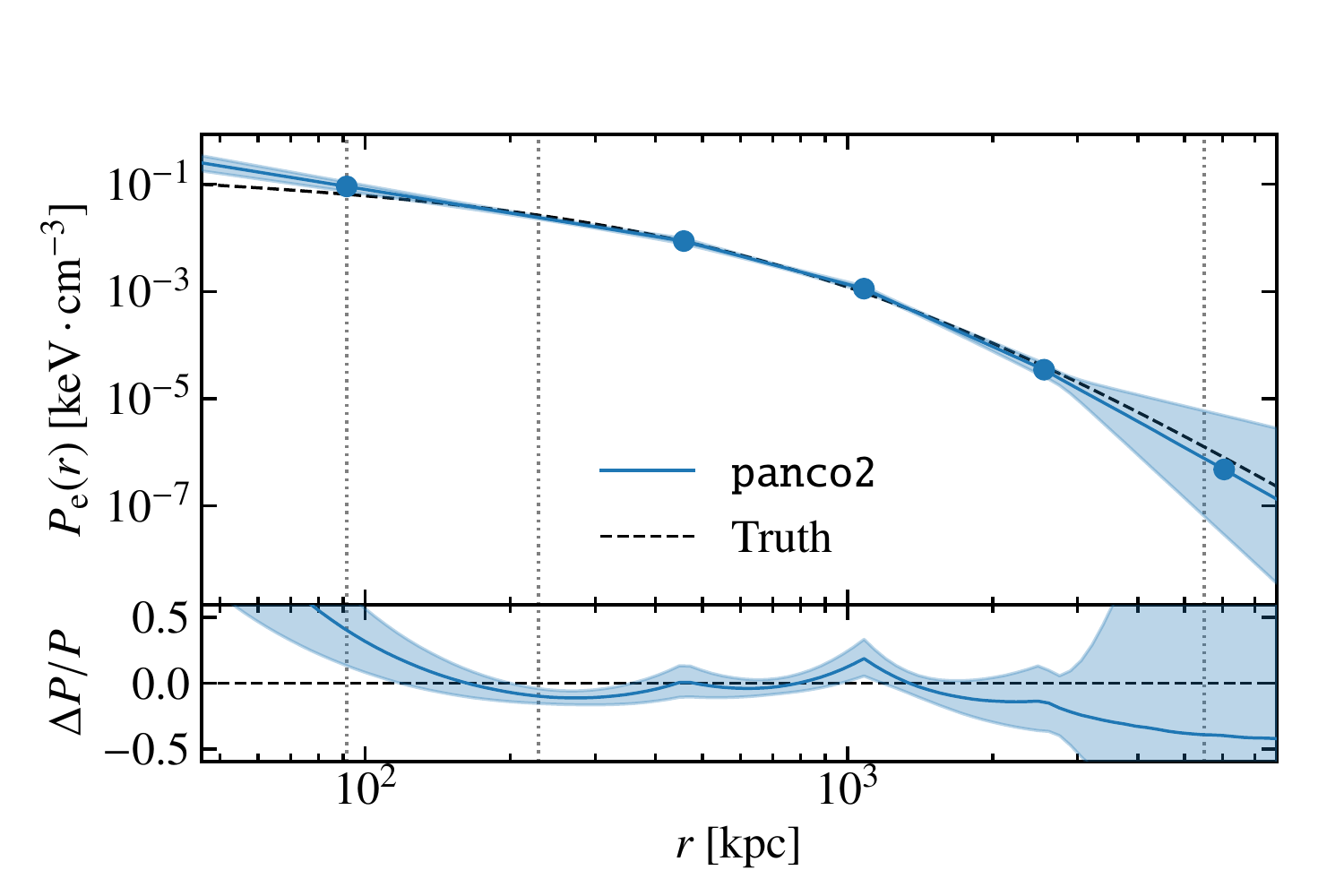} \\
    \rotatebox[x=10pt, y=55pt]{90}{C2, NIKA2}
    \includegraphics[height=4cm, trim={0 0 0 1cm}, clip]{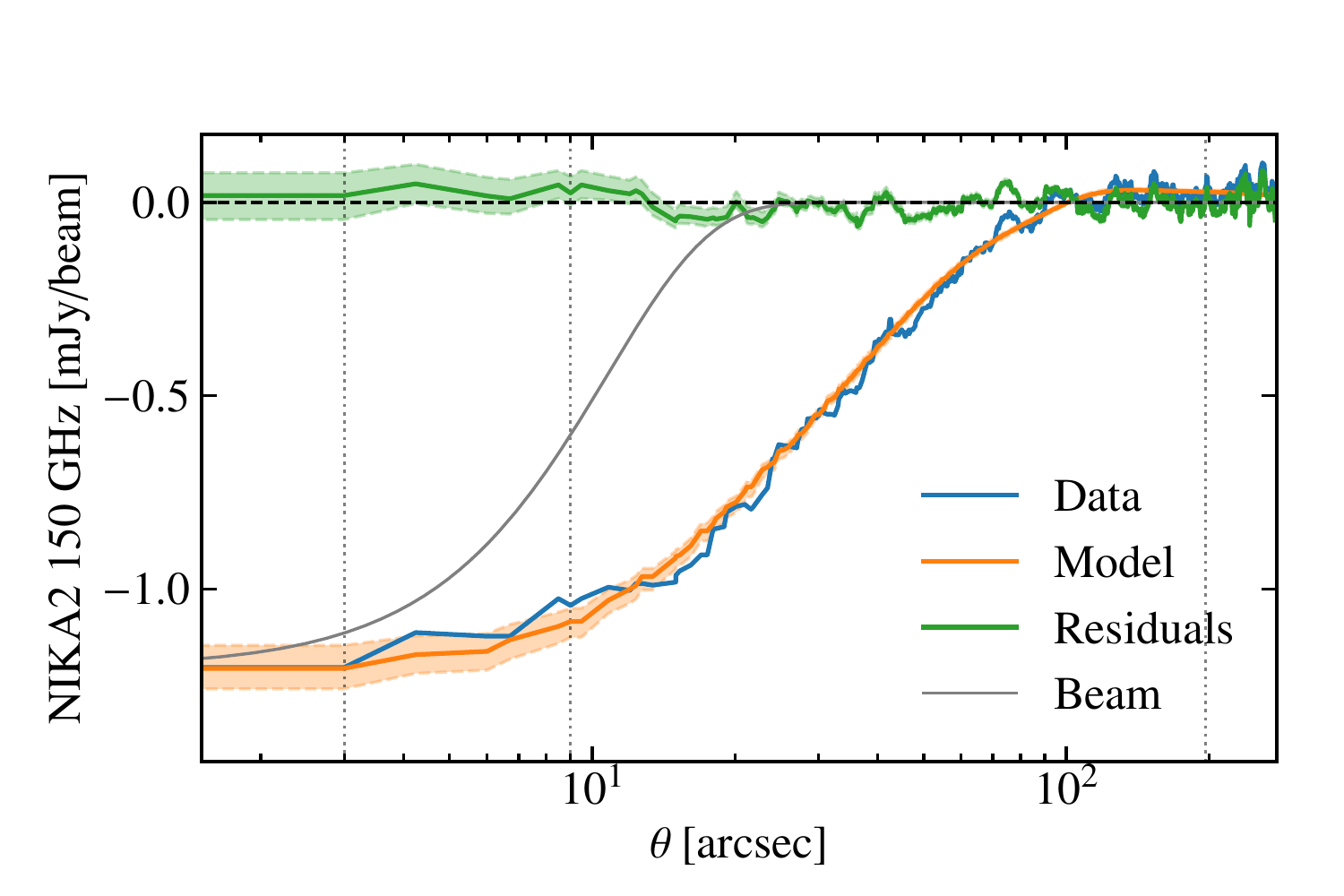}
    \includegraphics[height=4cm, trim={0 0 0 1cm}, clip]{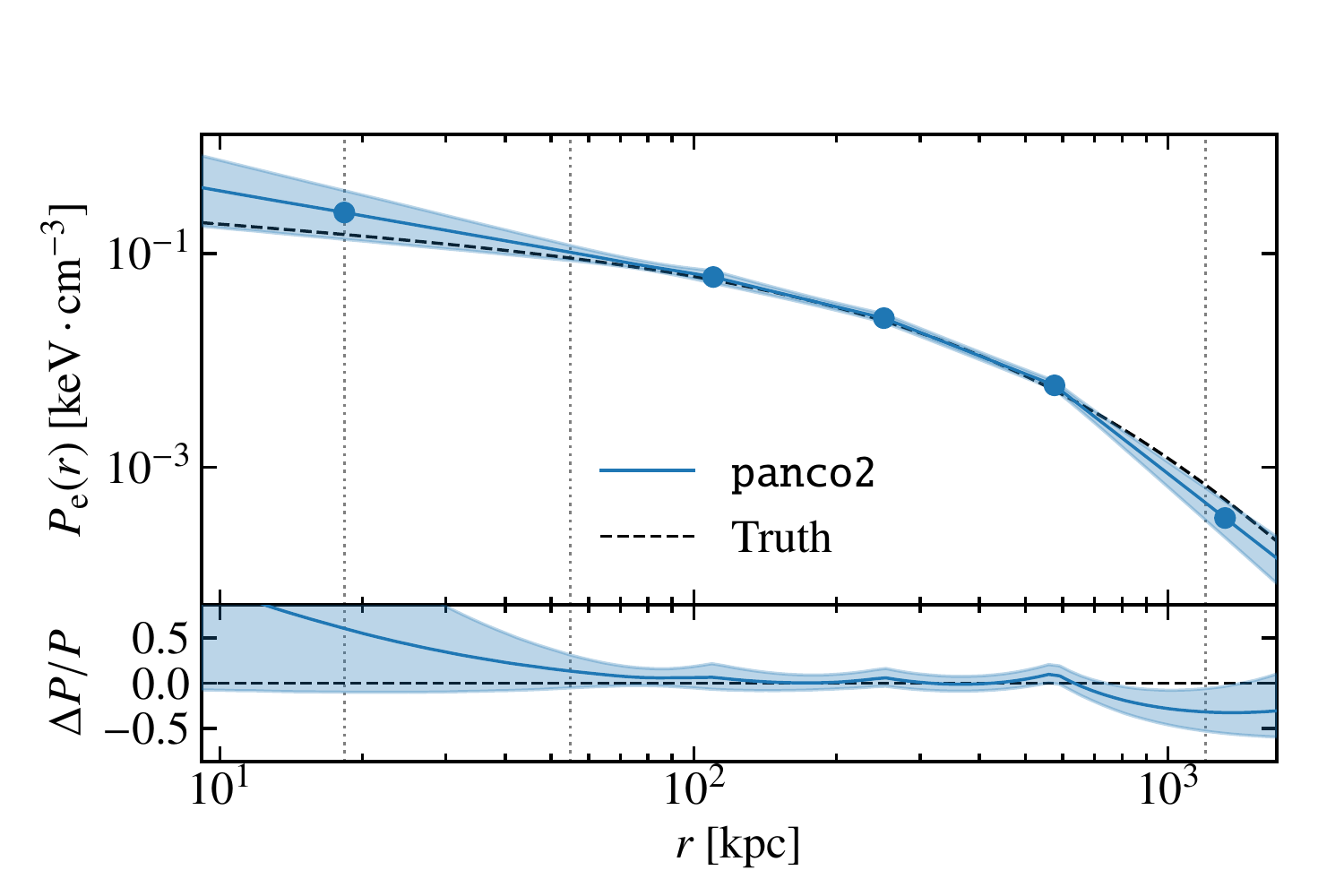} \\
    \rotatebox[x=10pt, y=55pt]{90}{C3, NIKA2}
    \includegraphics[height=4cm, trim={0 0 0 1cm}, clip]{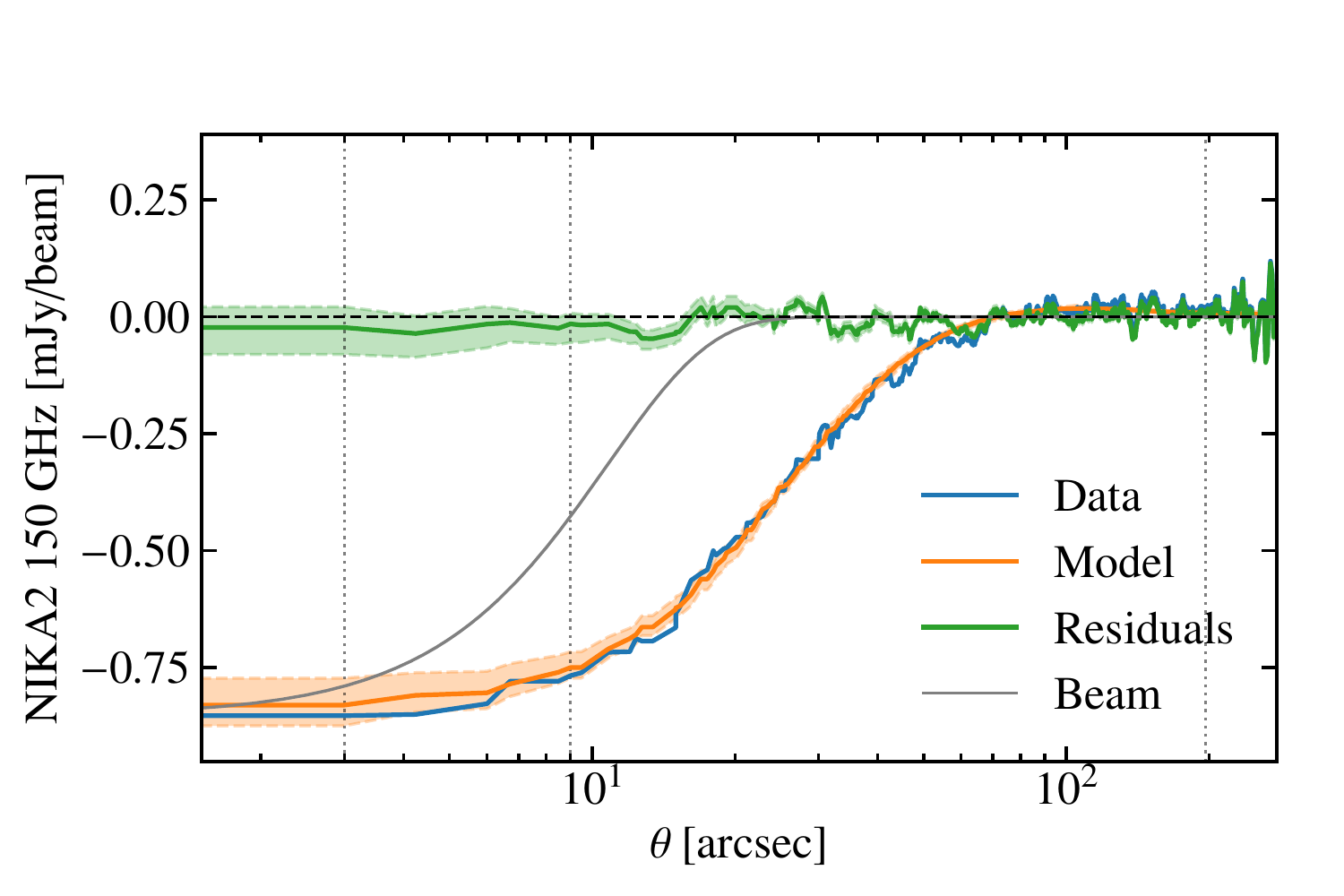}
    \includegraphics[height=4cm, trim={0 0 0 1cm}, clip]{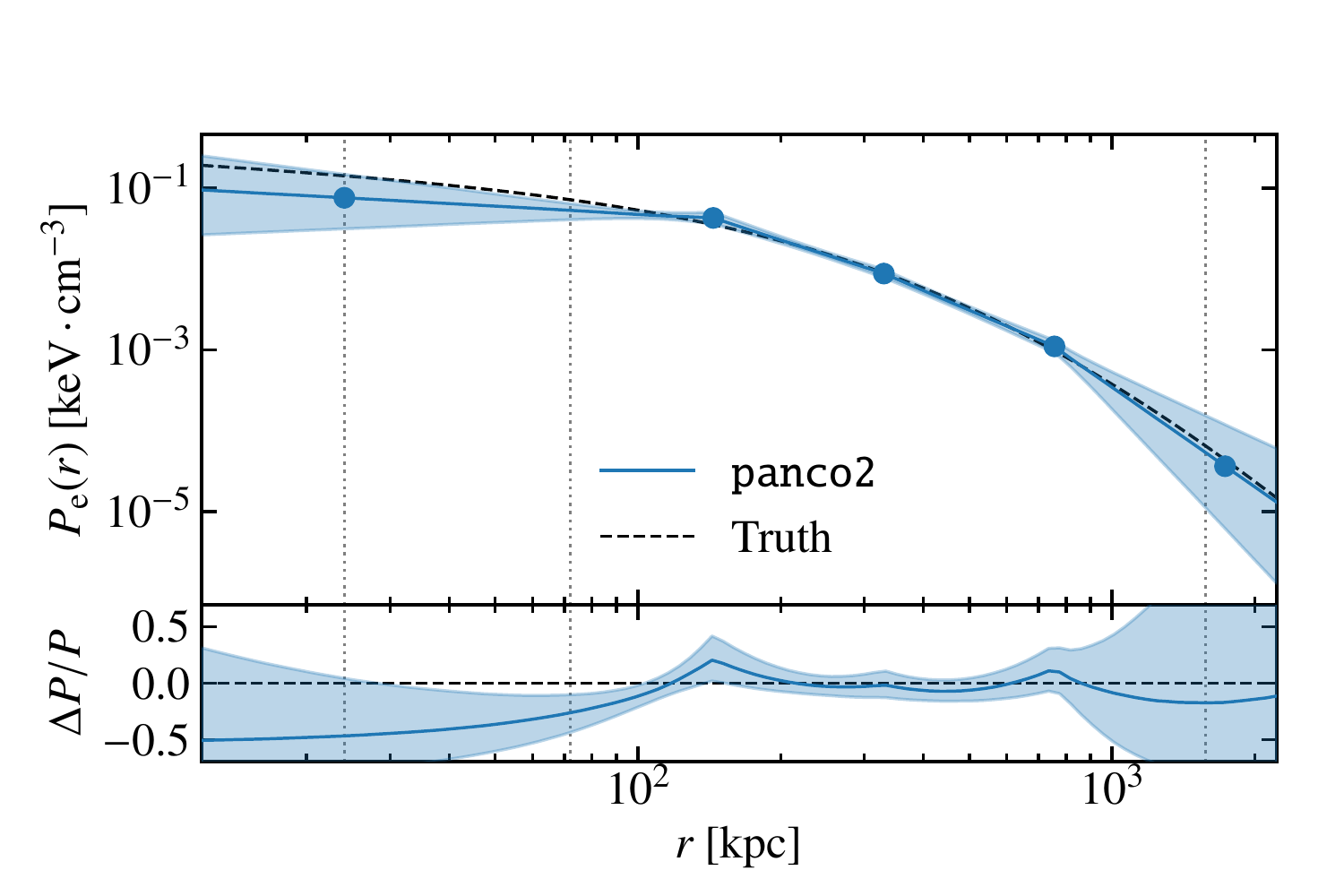}
    \caption{
        \textbf{Left:} Data, model, and residuals (\ie\ difference between data and model) azimuthal profiles for the five validation fits.
        The beam profile is shown in grey for comparison.
        The rows are identical to Figure~\ref{fig:valid:dmr_2d}.
        \textbf{Right:} Pressure profiles reconstructed for each validation fit (blue line).
        For each profile, the blue shaded area marks the region between the 16th and 84th percentiles of the posterior distribution.
        The dashed black line shows the true pressure profile used to generate each map.
        In each plot, dotted vertical lines, from left to right, show the size of a map pixel, the beam HWHM, and the half map size.
    }
    \label{fig:valid:profiles}
\end{figure*}

The results of the regression is presented in Figures~\ref{fig:valid:dmr_2d} and \ref{fig:valid:profiles}.
Figure~\ref{fig:valid:dmr_2d} shows, for each of the five datasets, the simulated data (left), best-fitting 2D model (center), and the residuals (right).
The residual maps contain no significant structure, providing a visual indication of goodness of fit.
The left panels of Figure~\ref{fig:valid:profiles} also shows the data, model, and residuals, but as 1D azimuthal profiles, and include uncertainty computed from the spread of the sampled posterior.
The compatibility of the green curves (\ie\ residuals) with zero within uncertainties proves that \panco\ is able to retrieve a pressure profile that fits the data.

To assess goodness of fit, we also compute reduced chi-squared values, defined as:
\begin{equation}
    \bar{\chi}^2 = -\frac{2}{k} \log \mathcal{L}(\vartheta_{\rm b.f.}),
\end{equation}
where the log-likelihood $\log \mathcal{L}$ is defined in \refeq{eq:algo:likelihood}, $k$ is the number of degrees of freedom, defined as the difference between the number of pixels in the input map and the number of dimensions in the parameter space, and $\vartheta_{\rm b.f.}$ is the parameter-space vector corresponding to the maximum likelihood sample in the Markov chains (\ie\ the ``best-fit'' vector). \\
Goodness of fit can be judged by a chi-squared value compatible with $1 \pm \sqrt{2k}/k$, corresponding to the spread of a reduced chi-squared probability distribution.
Four of the fits meet this criterion; the only marginal outlier is the fit of the SPT-like mapping of C1, for which the reduced chi-squared is 0.9\% too high to pass the test.

\begin{table}[t]
    \centering
    \begin{tabular}{cccc}
        \toprule
        Cluster & Instrument & $\bar{\chi}^2$ (d.o.f.) & Notes \\
        \midrule
        C1 & \textit{Planck} & 1.0098 (14634) & \S\ref{sec:simu} \\
        C1 & SPT & 1.0120 (14154) & \S\ref{sec:simu} \\
        C2 & SPT & 1.0067 (14154) & \S\ref{sec:simu} \\
        C2 & NIKA2 & 0.9932 (17154) & \S\ref{sec:simu} \\
        C3 & NIKA2 & 1.0027 (17154) & \S\ref{sec:simu} \\
        \midrule
        C2 & SPT & 1.0070 (14154) & Anisotropic filtering, \S\ref{sec:simu:2dtf} \\
        C2 & NIKA2 & 0.9932 (17152) & Point sources (fitted), \S\ref{sec:simu:ps} \\
        C2 & NIKA2 & 0.9962 (17154) & Point sources (masked), \S\ref{sec:simu:ps} \\
        C2 & NIKA2 & 0.9933 (17154) & $Y_{500}$ constraint, \S\ref{sec:simu:Ysz} \\
        \bottomrule
    \end{tabular}
    \caption{\normalfont
        Reduced chi-squared values corresponding to the best-fit parameters for each validation fit.
        We also include the number of freedom in the fit, defined as the difference between the number of pixels in the map and the number of parameters in the model.
    }
    \label{tab:simu:chi2}
\end{table}

The right panels of Figure~\ref{fig:valid:profiles} show the pressure profiles recovered for each fit as the blue curves, including uncertainties.
The true profile used to generate each map data (\ie\ the universal \aten\ pressure profile for the cluster's mass and redshift, see \S\ref{sec:simu:mkdata}) is shown as a black dashed line.
For each data set, the agreement between the two curves across the considered radial range (\ie\ from the projected instrumental resolution to the projected half map size) shows that \panco\ is able to retrieve an accurate estimation of the pressure profile of a cluster from its tSZ map.
As discussed in \S\ref{sec:simu:fit}, even though a radial bin is included in that range, we advise treading lightly when drawing conclusions from pressure profile fits at radii smaller than the projected resolution of the map, \ie\ the PSF of the instrument used to perform the tSZ mapping.

Finally, posterior surfaces for the five fits are presented and discussed in Appendix~\ref{sec:ap:corner}, showing the sampled posterior in the parameter space, and the corresponding reduced chi-squared distributions.

\section{Additional simulated datasets} \label{sec:extra_simu}

The simulations described in \S\ref{sec:simu} did not feature all the possibilities of pressure profile fitting offered by \panco.
To extend the validation to these different analysis options, we create additional datasets, based on the same mock clusters and instrumental data coverages, but each featuring one more component to the modeling.

\begin{figure}[t]
    \centering
    \includegraphics[width=0.95\linewidth, trim={0 0 0 1cm}, clip]{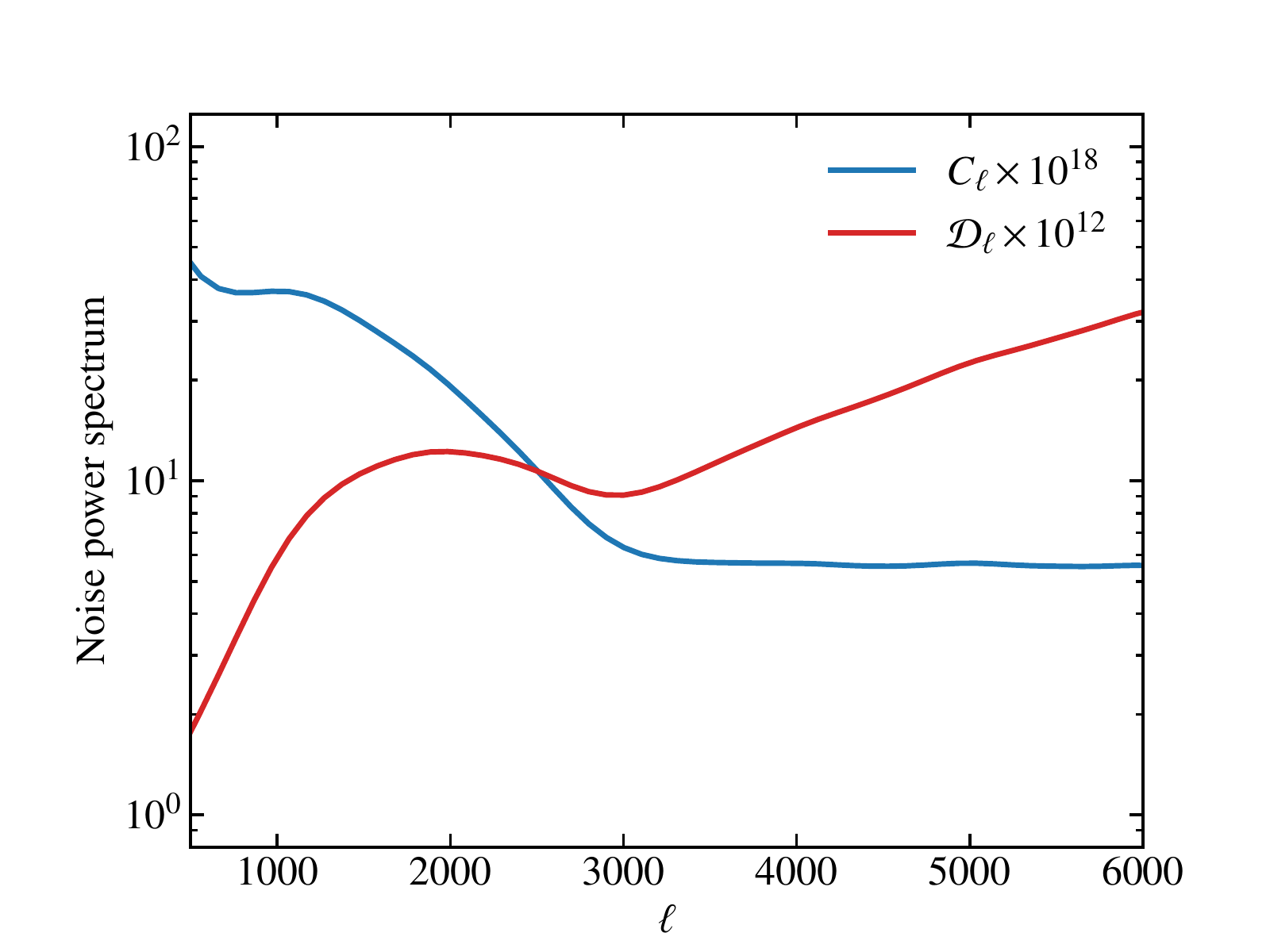}
    \includegraphics[width=0.95\linewidth, trim={0 0 0 1cm}, clip]{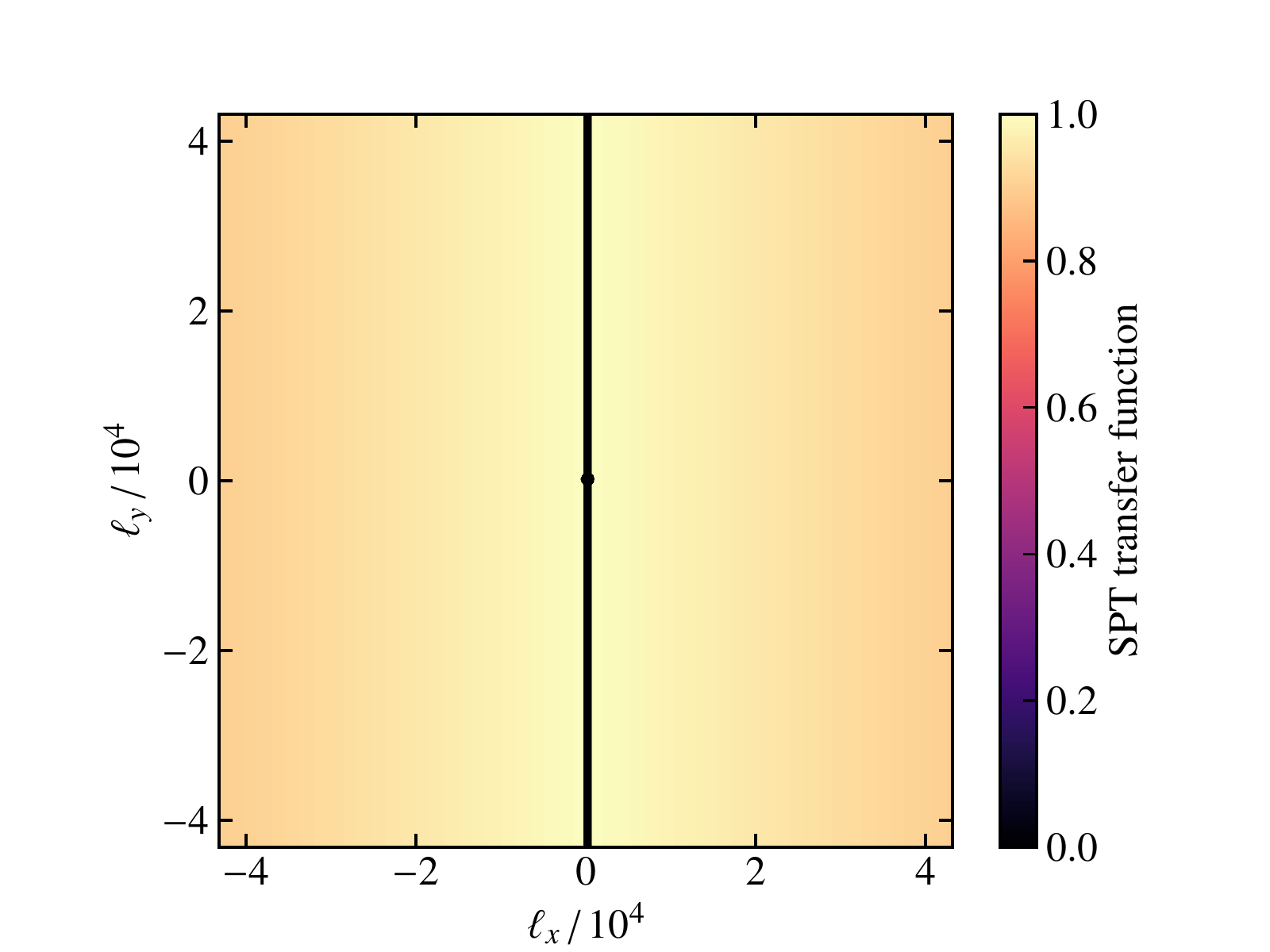}
    \caption{
        Additional features added to SPT-like maps for the validation of \panco.
        \textbf{Top:} Power spectrum of the added correlated noise in \S\ref{sec:simu:corr_noise}, estimated from a $(5\degree \times 5\degree)$ patch of the minimum-variance SPT $y-$map of \sptymap\ and smoothed.
        We also show $\mathcal{D}_\ell \equiv \ell (\ell + 1) C_\ell / 2\pi$ in red for illustrative purposes.
        \textbf{Bottom:} Anisotropic transfer function used in \S\ref{sec:simu:2dtf}.
        The filtering is artificially created to mimic single-band SPT transfer functions \citep[see \eg\ Figure~8 of][]{schaffer_first_2011}.
    }
    \label{fig:valid2:spt_complications}
\end{figure}

\subsection{Spatially correlated noise} \label{sec:simu:corr_noise}

We create a mock SPT map of the C2 galaxy cluster featuring realistic correlated noise.
We evaluate the power spectrum of noise in the SPT $y-$map from a $(5\degree \times 5\degree)$ patch of the minimum variance map published in \sptymap, in which astrophysical sources were masked.
The resulting power spectrum is smoothed, and illustrated in the top panel of Figure~\ref{fig:valid2:spt_complications}.
It shows significant noise power at large angular scales, and a relatively flat noise at $\ell > 3000$.
For illustration purposes, we also show $\mathcal{D}_\ell$, which can be compared to the power spectrum of the full SPT $y-$map in \sptymap.
We use this power spectrum to create $10^4$ random noise realizations, that are used to compute the covariance matrix of the noise in map pixels.
One of these noise realizations is added to the simulated tSZ signal, and the inverted noise covariance is used in the likelihood function of eq.~(\ref{eq:algo:likelihood}).
We note that the MCMC sampling is substantially slower in the presence of a noise covariance matrix, due to the high-dimensionality of the matrix product that needs to be computed to evaluate the likelihood.
To make the process quicker, we choose to focus on a $(20' \times 20')$ map for this analysis, covering a sky patch smaller than those we have considered for SPT maps.

\begin{figure}[t]
    \centering
    \includegraphics[width=0.95\linewidth, trim={0 0 0 1cm}, clip]{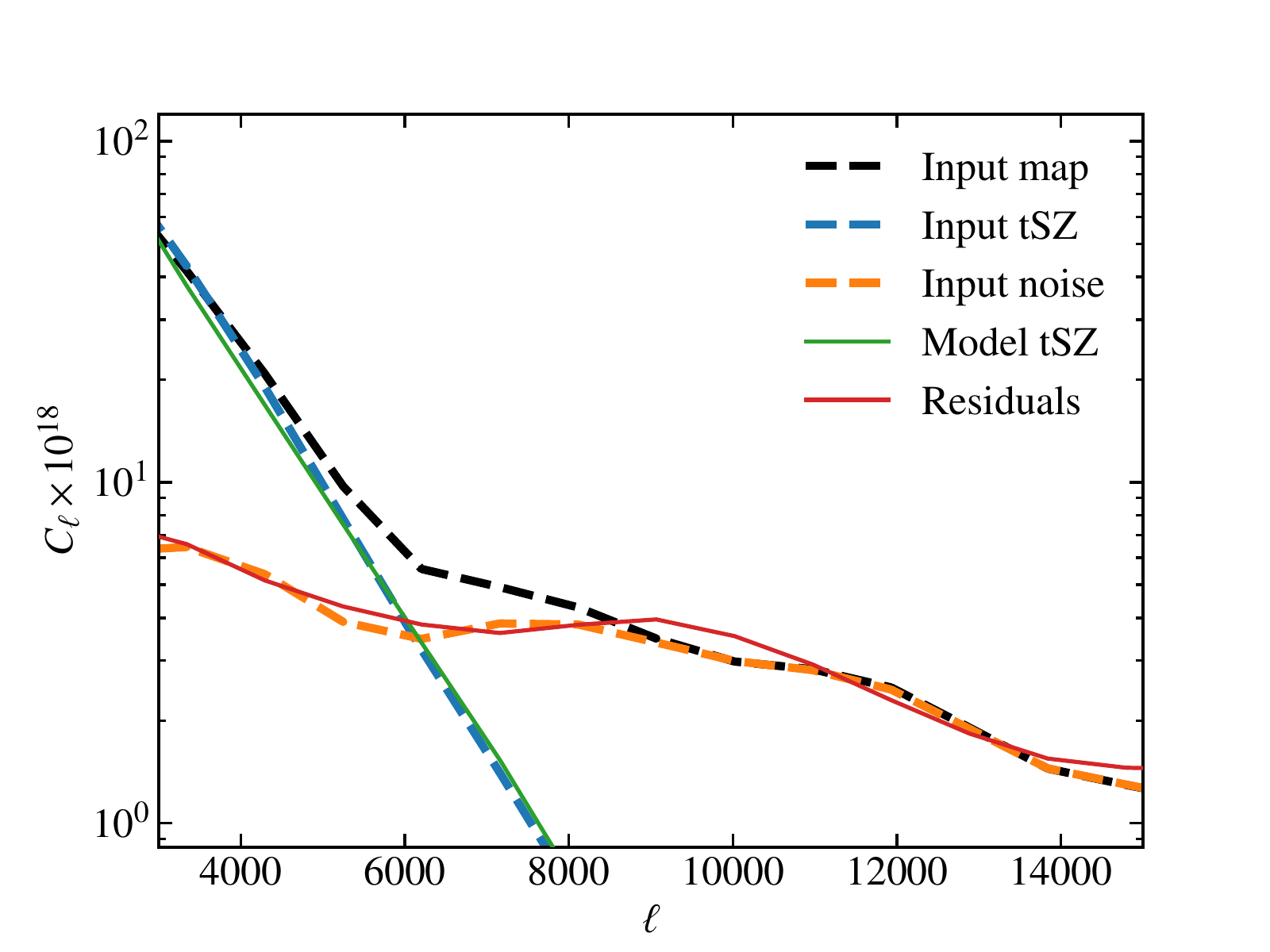}
    \includegraphics[width=0.95\linewidth, trim={0 0 0 1cm}, clip]{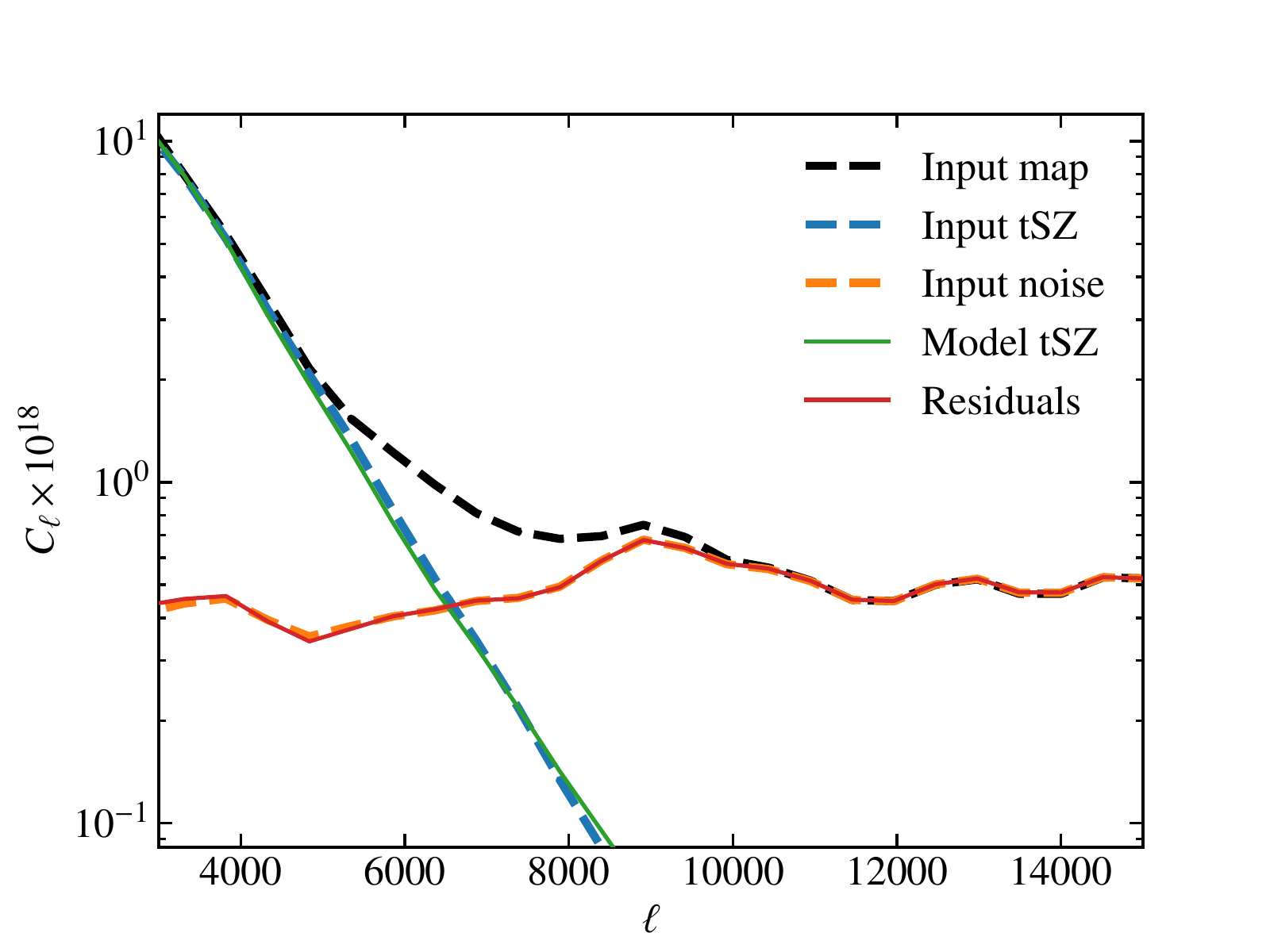}
    \caption{
        Angular power spectra of the different components in the SPT-like maps of C2 including correlated noise (\textit{top}) and an anisotropic filtering (\textit{bottom}).
        The total power spectrum (black) is the sum of that of the tSZ signal (blue) and of the noise (orange).
        The green and orange lines respectively show the power spectrum of the model map computed from the median of the posterior sampled by \panco, and of the corresponding residuals.
        The agreement between the blue and green lines (or, equivalently, between the red and orange lines) shows that the model reconstructed by \panco\ correctly reproduces the tSZ signal in Fourier space, even in the presence of colored noise and complex large-scale filtering.
    }
    \label{fig:valid2:spt_complications_powspec}
\end{figure}

\begin{figure*}[t]
    \centering
    \includegraphics[height=4cm, trim={1.5cm 0cm 1.5cm 1.75cm}, clip]{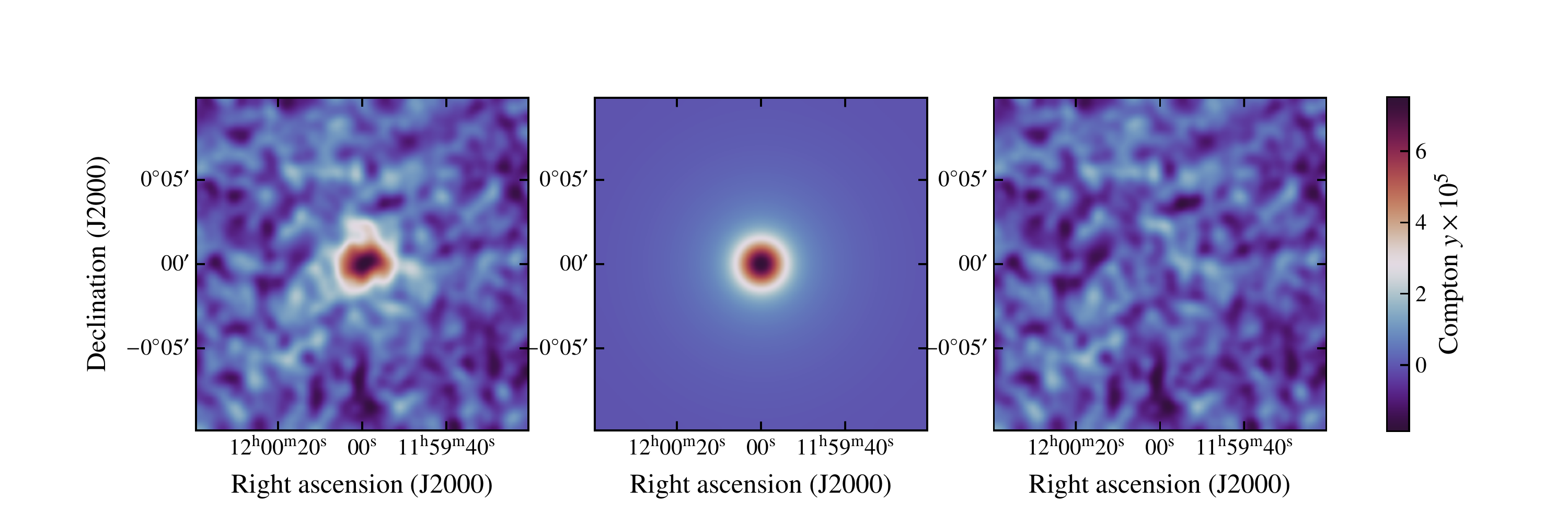}
    \includegraphics[height=4cm, trim={0 0 0 1cm}, clip]{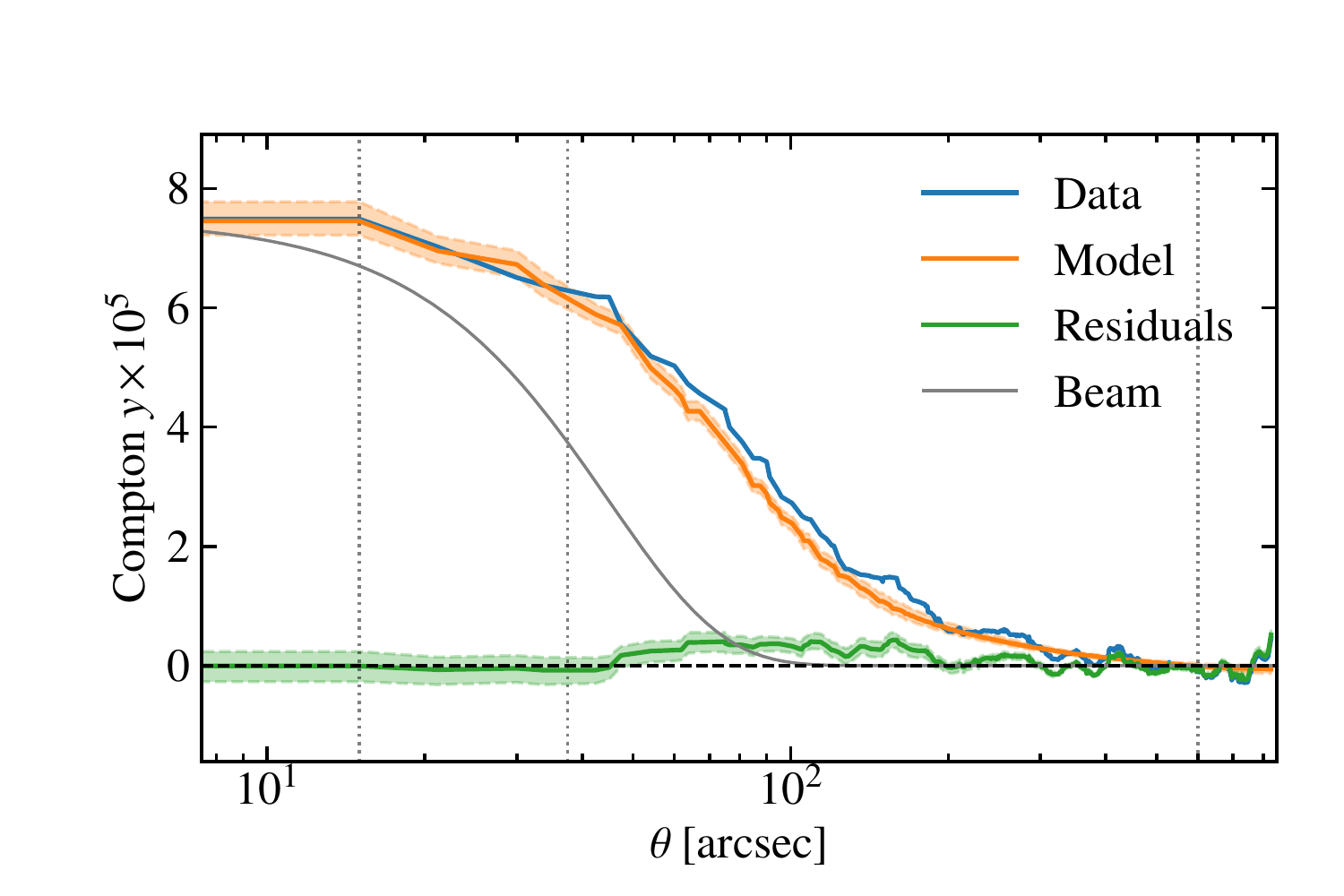}
    \includegraphics[height=4cm, trim={0 0 0 1cm}, clip]{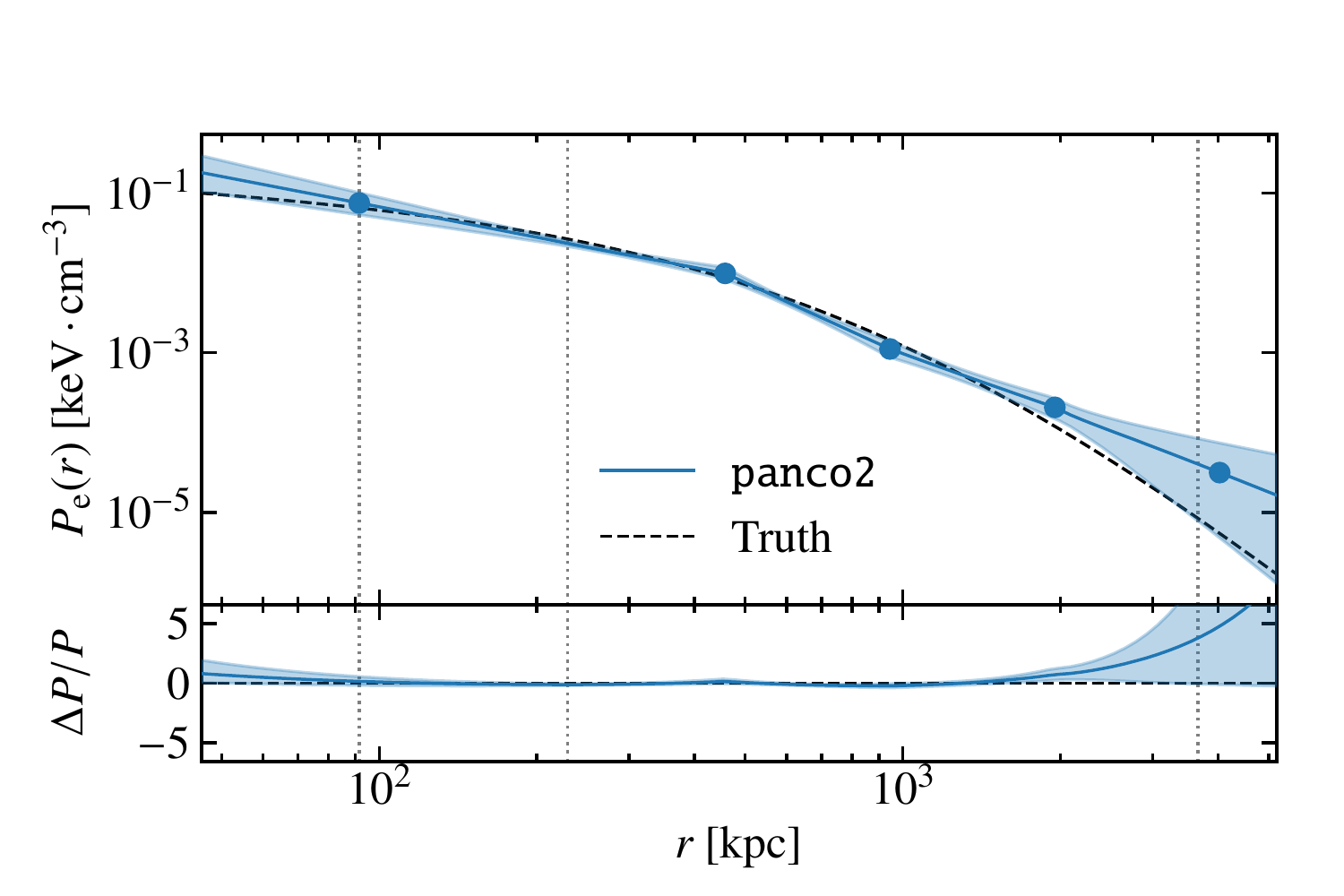}
    \caption{
        Validation results for the C2 cluster seen by SPT, including correlated noise (see \S\ref{sec:simu:corr_noise}).
        \textbf{Top:} Data (\textit{left}), model (\textit{center}), and residuals (\ie\ difference between data and model, \textit{right}).
        Maps are smoothed by a $1$ pixel Gaussian for display purposes.
        \textbf{Bottom:} Radial profiles for the data, model and residuals maps (\textit{left}) and recovered pressure profile (\textit{right}).
        Error envelopes the region between the 16th and 84th percentiles of the posterior distribution.
        Dotted vertical lines, from left to right, show the size of a map pixel, the beam HWHM, and the half map size.
        The dashed black line shows the true pressure profile used to generate each map.
    }
    \label{fig:valid2:corrnoise}
\end{figure*}

The pressure profile fitting is performed the same way as described in \S\ref{sec:simu:fit}.
The results are presented\footnote{We do not report on the value of the reduced chi-squared for this fit, as it is only a relevant statistic when pixels are independant, which is not the case when the likelihood includes a non-diagonal covariance matrix.} in Figure~\ref{fig:valid2:corrnoise}.
Large-scale correlated noise in the map proves to be a challenge in the pressure reconstruction, as the recovered pressure profile is overestimated in the outskirts (at $r \geqslant 2 \, {\rm Mpc}$).
Nonetheless, it is still compatible with the truth within $1\sigma$ uncertainties.
We also show the results in Fourier space in the top panel of Figure~\ref{fig:valid2:spt_complications_powspec}.
We see that despite the injected noise being colored (\ie\ its power spectrum not being flat), the inclusion of the covariance matrix in the likelihood allows \panco\ to recover a model map with a power spectrum very close to that of the input tSZ signal, and that the residuals are compatible with noise in their structure in Fourier space.

\subsection{Two-dimensional transfer function} \label{sec:simu:2dtf}

We create a mock SPT map of the C2 galaxy cluster featuring anisotropic filtering by a two-dimensional transfer function.
Even though the published SPT $y-$maps have negligible filtering (see \sptymap), single-band SPT maps used for cluster finding \citep[and cluster science, see \eg][]{young_mean_nodate} are created with a high-pass filter in $\ell_x$ corresponding to filtering along the right ascension axis, corresponding to the scanning strategy direction \citep{schaffer_first_2011}.
We create a mock map featuring an artificial anisotropic filtering mimicking this transfer function, which is accounted for in the forward modeling of the tSZ map.
Specifically, our mock transfer function is made of a high-pass filter in $\ell$ ($\ell > 800$) and in $\ell_x$ ($\ell_x > 500$).
We also add a slight descending gradient with increasing $|\ell_x|$.
The resulting transfer function is shown in the bottom panel of Figure~\ref{fig:valid2:spt_complications}.
We emphasize that this transfer function is not characteristic to the actual published SPT $y-$map, but rather an arbitrary tool that we chose to assess the ability of \panco\ to recover accurate pressure measurements in the case of an anisotropic filtering.

\begin{figure*}[t]
    \centering
    \includegraphics[height=4cm, trim={1.5cm 0cm 1.5cm 1.75cm}, clip]{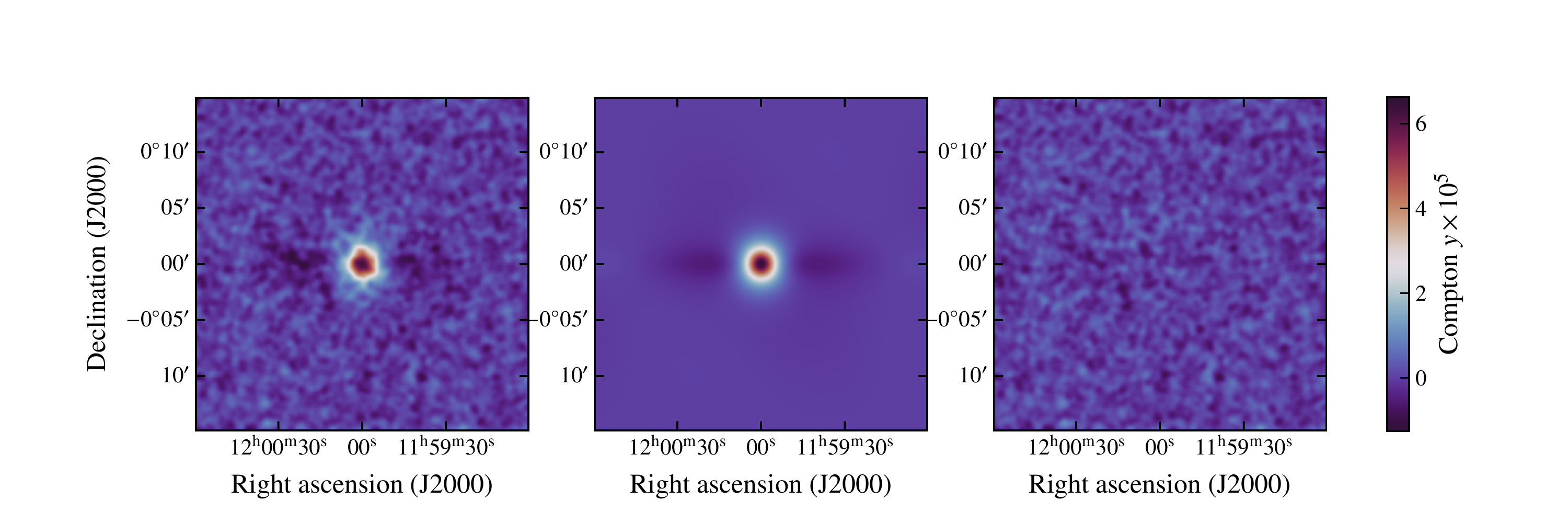}
    \includegraphics[height=4cm, trim={0 0 0 1cm}, clip]{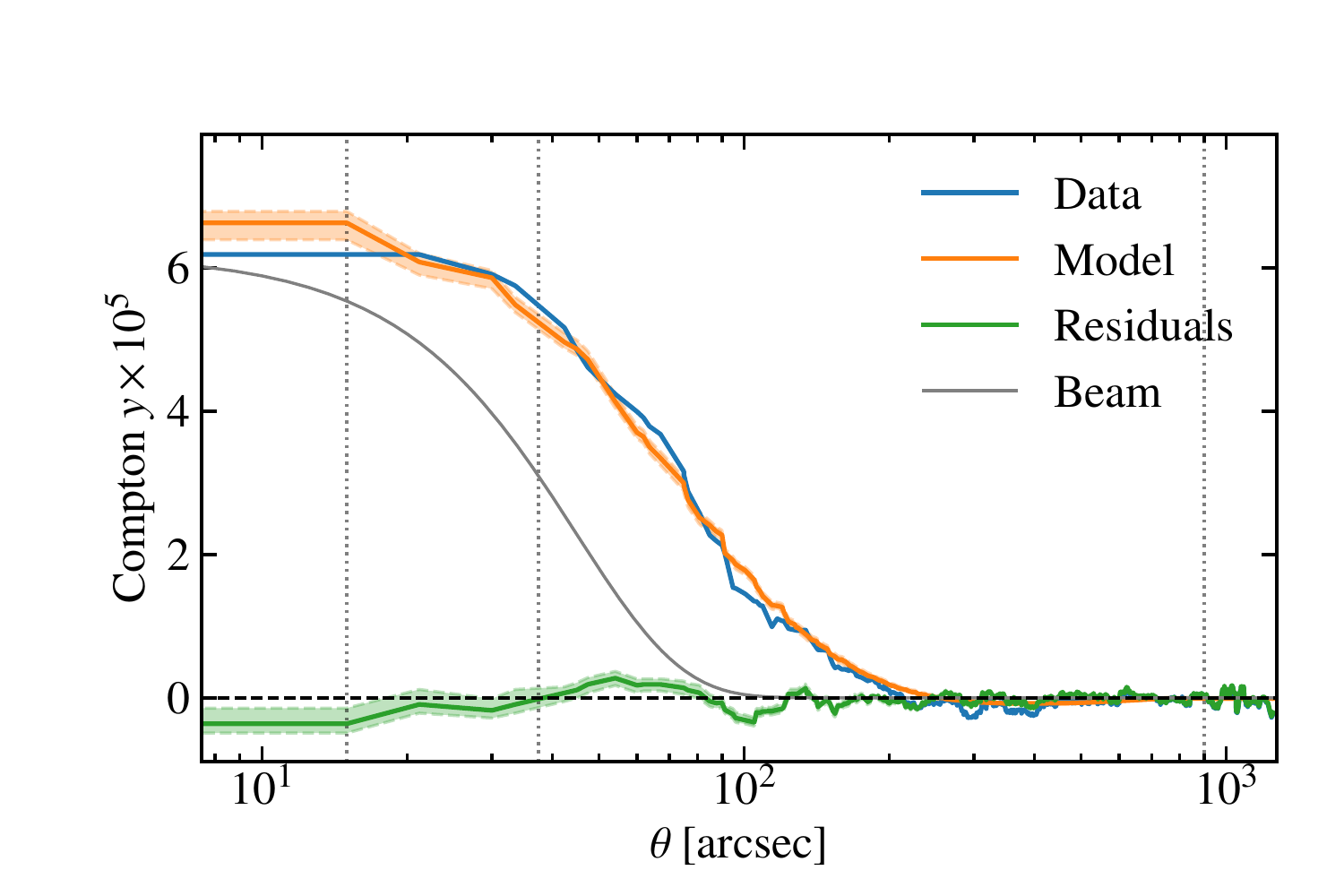}
    \includegraphics[height=4cm, trim={0 0 0 1cm}, clip]{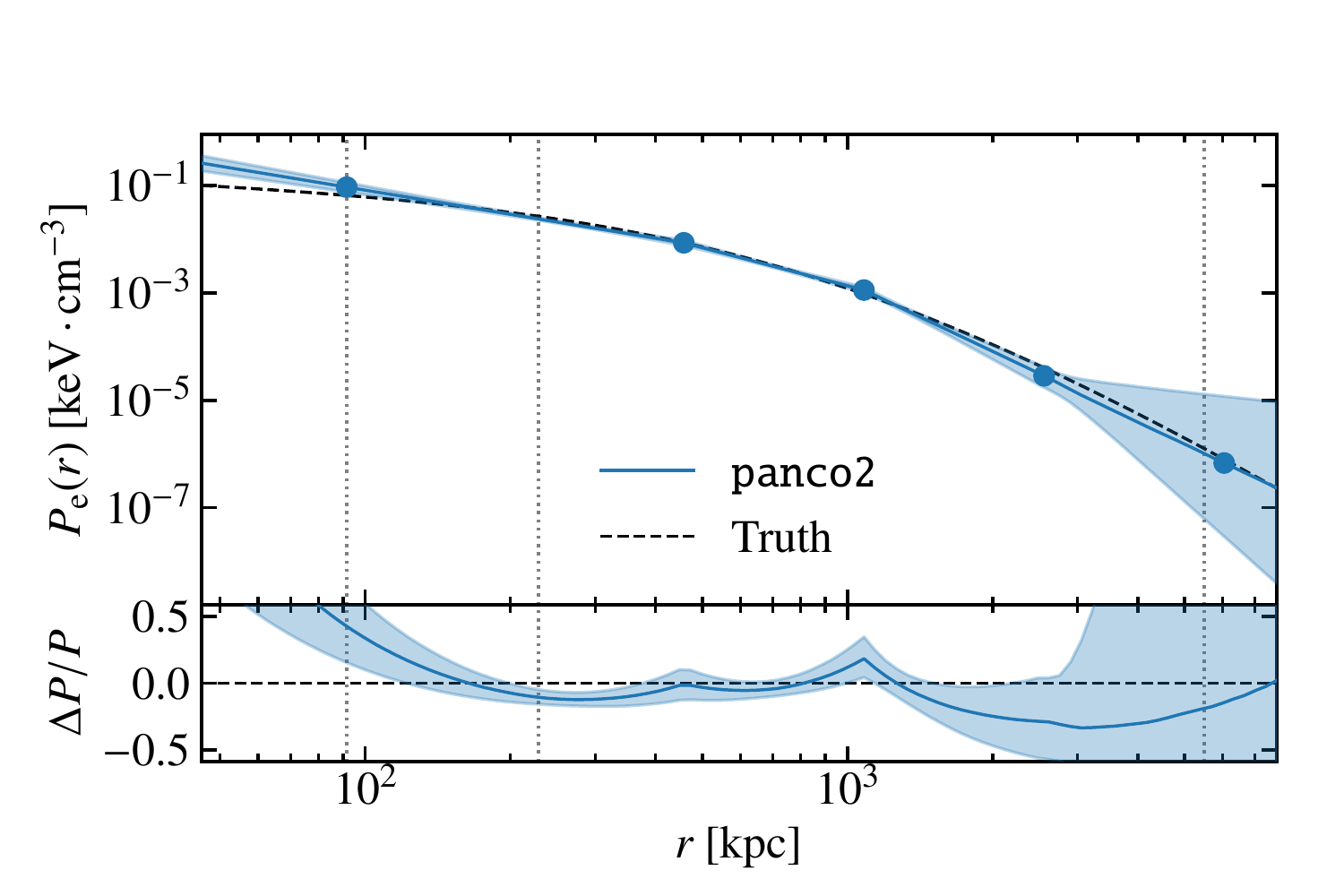}
    \caption{
        Validation results for the C2 cluster seen by SPT, including an anisotropic filtering (see \S\ref{sec:simu:2dtf}).
        Legend is identical to Fig.~\ref{fig:valid2:corrnoise}.
    }
    \label{fig:valid2:tf2d}
\end{figure*}

Results are shown in Figure \ref{fig:valid2:tf2d}.
The input map shows wing-like structures along the $x-$axis of the map, characteristic of a high-pass filter in $\ell_x$ \citep[see \eg][]{schaffer_first_2011}.
These are reproduced in the model computation, and the residuals do not show evidence of significant badness of fit.
The pressure profile is recovered within uncertainties from the pixel size to the map size, attesting that the anisotropic filtering has been accurately taken into account.
We also show the results in Fourier space in the bottom panel of Figure~\ref{fig:valid2:spt_complications_powspec}.
The power spectrum of the residuals is nearly indistinguishable from that of the noise, showing that the anisotropic filtering in the input data does not prevent the recovery of a tSZ model map, as long as this filtering is known and accounted for.
In addition, the best-fitting reduced chi-squared reported in table~\ref{tab:simu:chi2} also indicates goodness of fit.

\subsection{Point source contamination and map masking} \label{sec:simu:ps}

We generate a mock NIKA2 map of the C2 cluster in which point sources are added to the surface brightness map.
We choose to add two point sources S1 and S2, each with a flux of $F = 1 \, {\rm mJy}$, and respectively located at $30''$ and $75''$ from the cluster center.
They are added to the sky model with their true positions, and fluxes are treated as a model parameter as described in \S\ref{sec:algo:fwdmod}.

\begin{figure*}[t]
    \centering
    \includegraphics[height=4cm, trim={1.5cm 0cm 1.5cm 1.75cm}, clip]{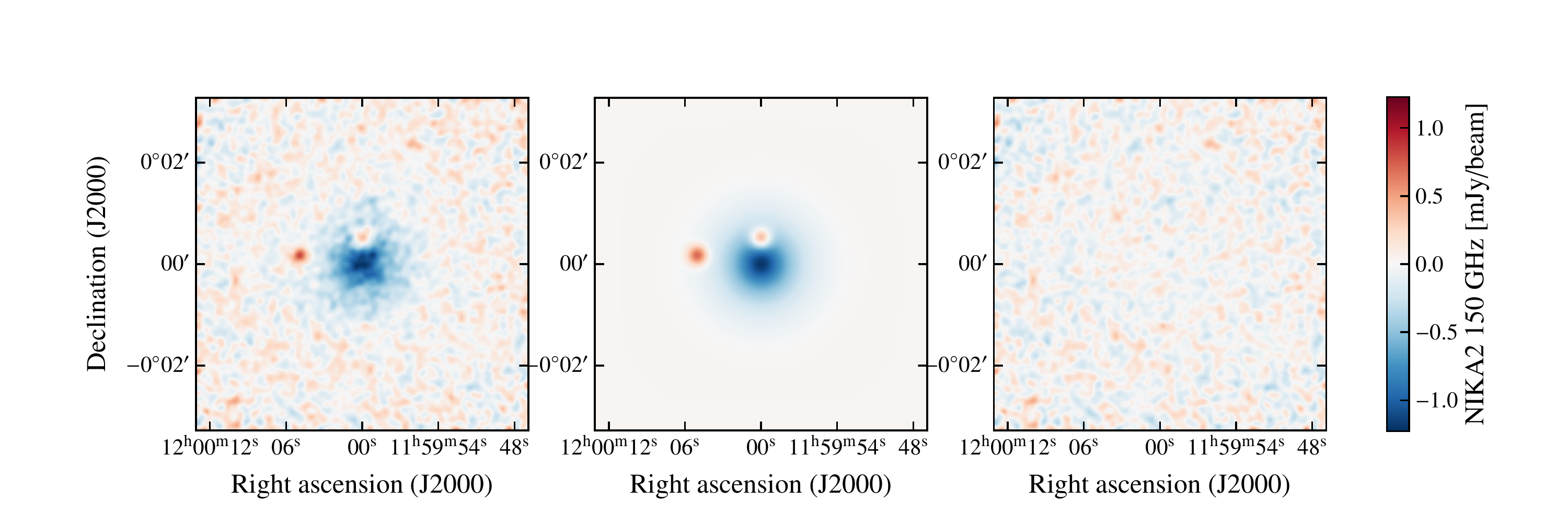}
    \includegraphics[height=4cm, trim={0 0 0 1cm}, clip]{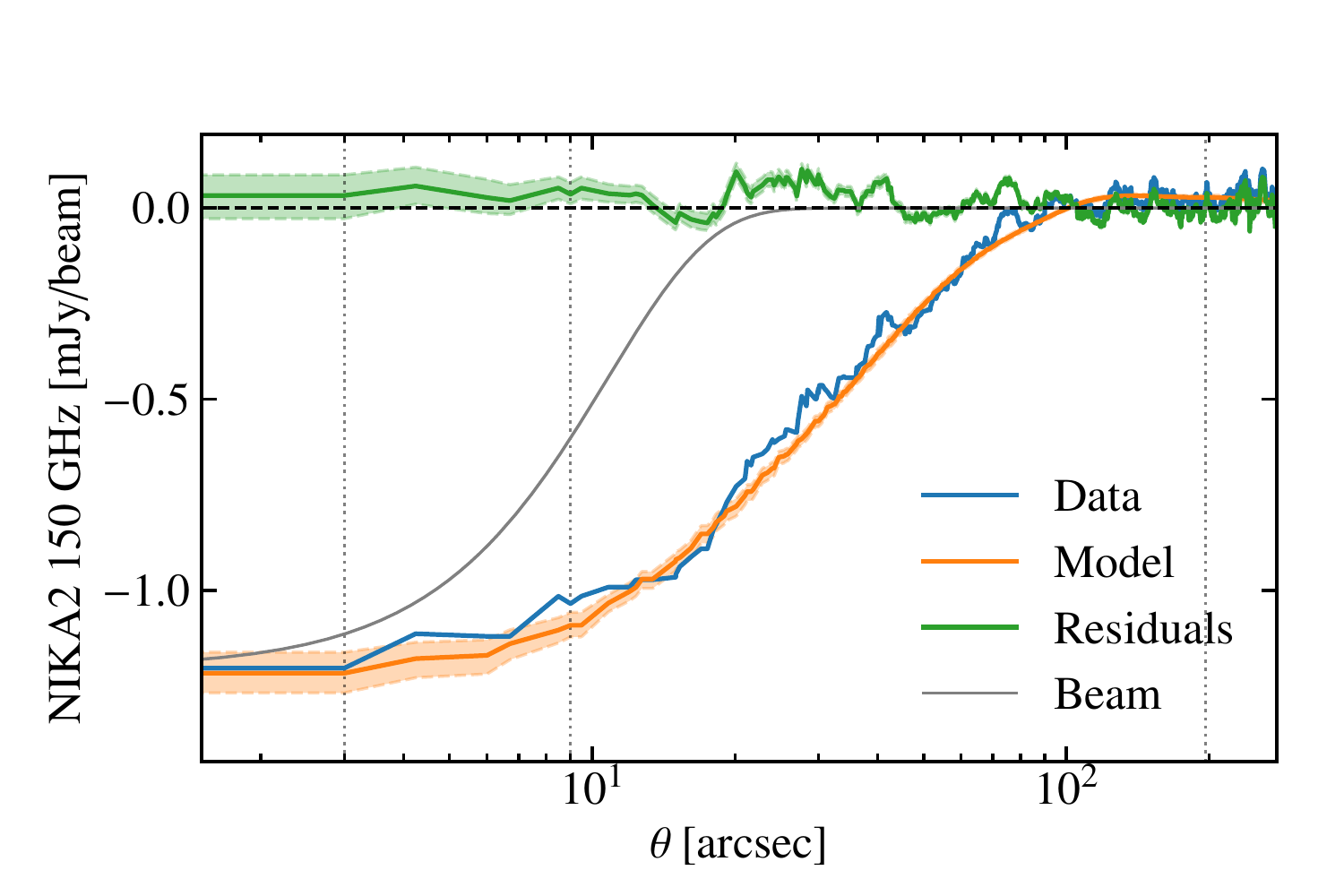}
    \includegraphics[height=4cm, trim={0 0 0 1cm}, clip]{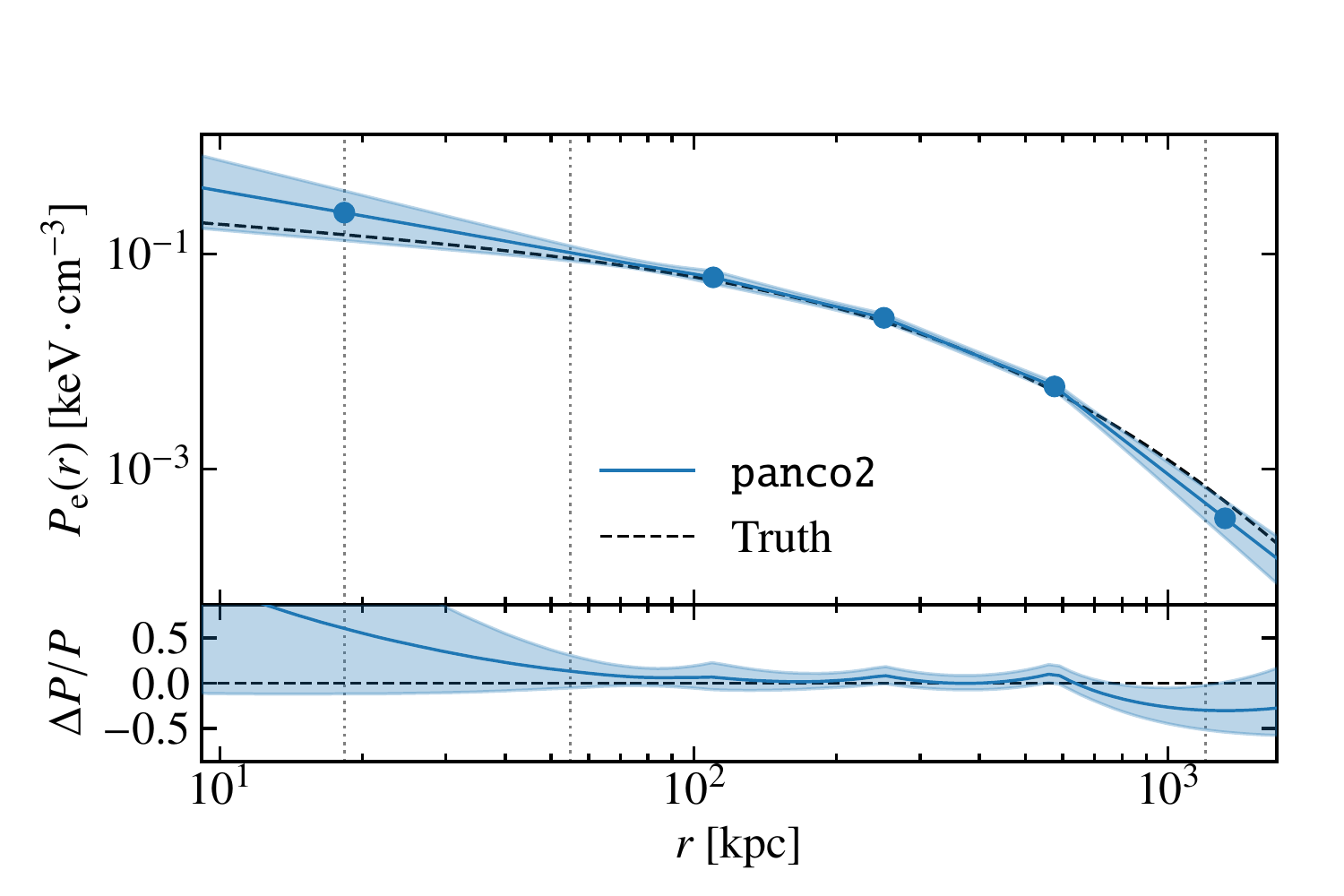}
    \caption{
        Validation results for the C2 cluster seen by NIKA2, including point source contamination (see \S\ref{sec:simu:ps}).
        Legend is identical to Fig.~\ref{fig:valid2:corrnoise}.
        The model map is the sum of tSZ surface brightness and of the contribution by point sources.
    }
    \label{fig:valid2:ps}
\end{figure*}

The fitting is performed the same way as described in \S\ref{sec:simu:fit}.
Priors on their fluxes are added as a Gaussian distribution for S1, with $p(F_1) = \mathcal{N}(1 \,{\rm mJy}, 0.2 \,{\rm mJy})$, and as an uniform distribution for S2, corresponding to an upper limit on the flux, with $p(F_2) = \mathcal{U}(0 \,{\rm mJy}, 2 \,{\rm mJy})$.
The results are presented in Figure \ref{fig:valid2:ps}.
Residuals do not include significant signal, either associated to tSZ signal or to point sources.
The pressure profile is accurately recovered within uncertainties on the radial range covered by the data.

Alternatively, if little is known about the point sources and informative priors on their fluxes cannot be provided, they can be masked from the analysis.
We repeat the fit of the same pressure profile model in the same map, but without adding point sources to the map model.
Instead, we mask a circular patch with diameter the FWHM of the NIKA2 beam at the location of each point source.
For scales larger than the instrumental beam, the pressure profiles estimated by masking the point sources and by modelling them are nearly identical, both being able to recover the true profile.
We note that the masked analysis yields a profile estimate too high in the central region of the cluster, which can be interpreted as due to a lack of constraining data, as one of the sources is very close to the center, and masking it removes an important fraction of the constraining data.

For both methods, the reduced chi-squared is reported in table~\ref{tab:simu:chi2}, attesting of goodness of fit.

\begin{figure*}[t]
    \centering
    \includegraphics[height=4cm, trim={1.5cm 0cm 1.5cm 1.75cm}, clip]{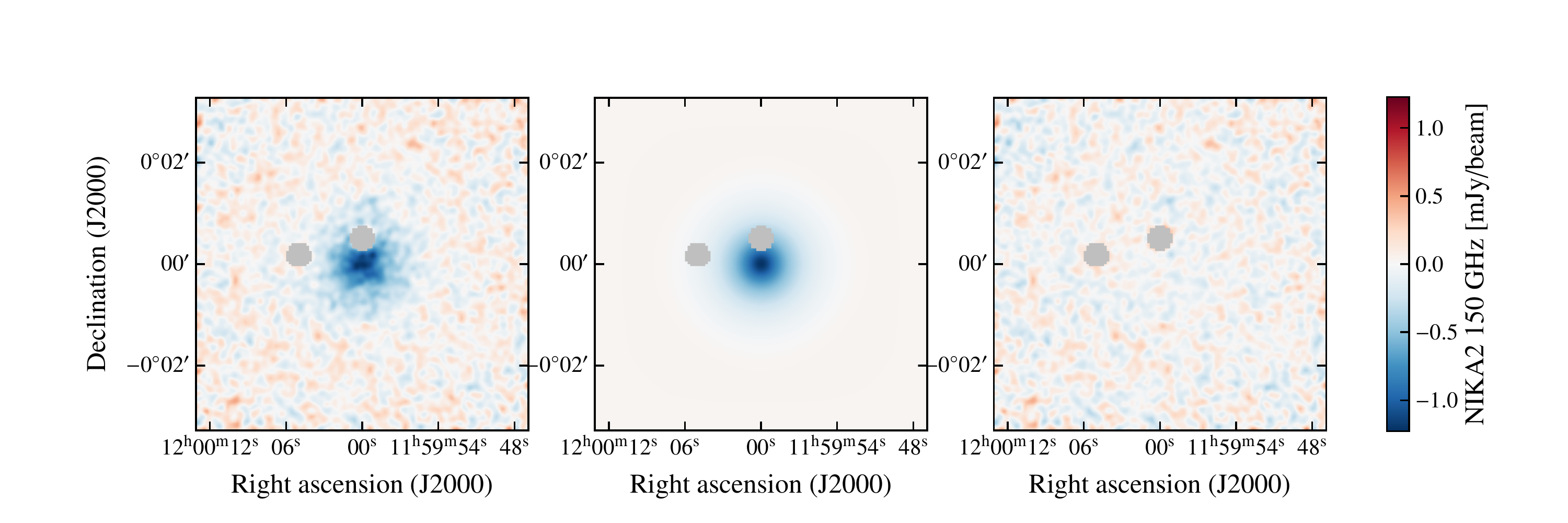}
    \includegraphics[height=4cm, trim={0 0 0 1cm}, clip]{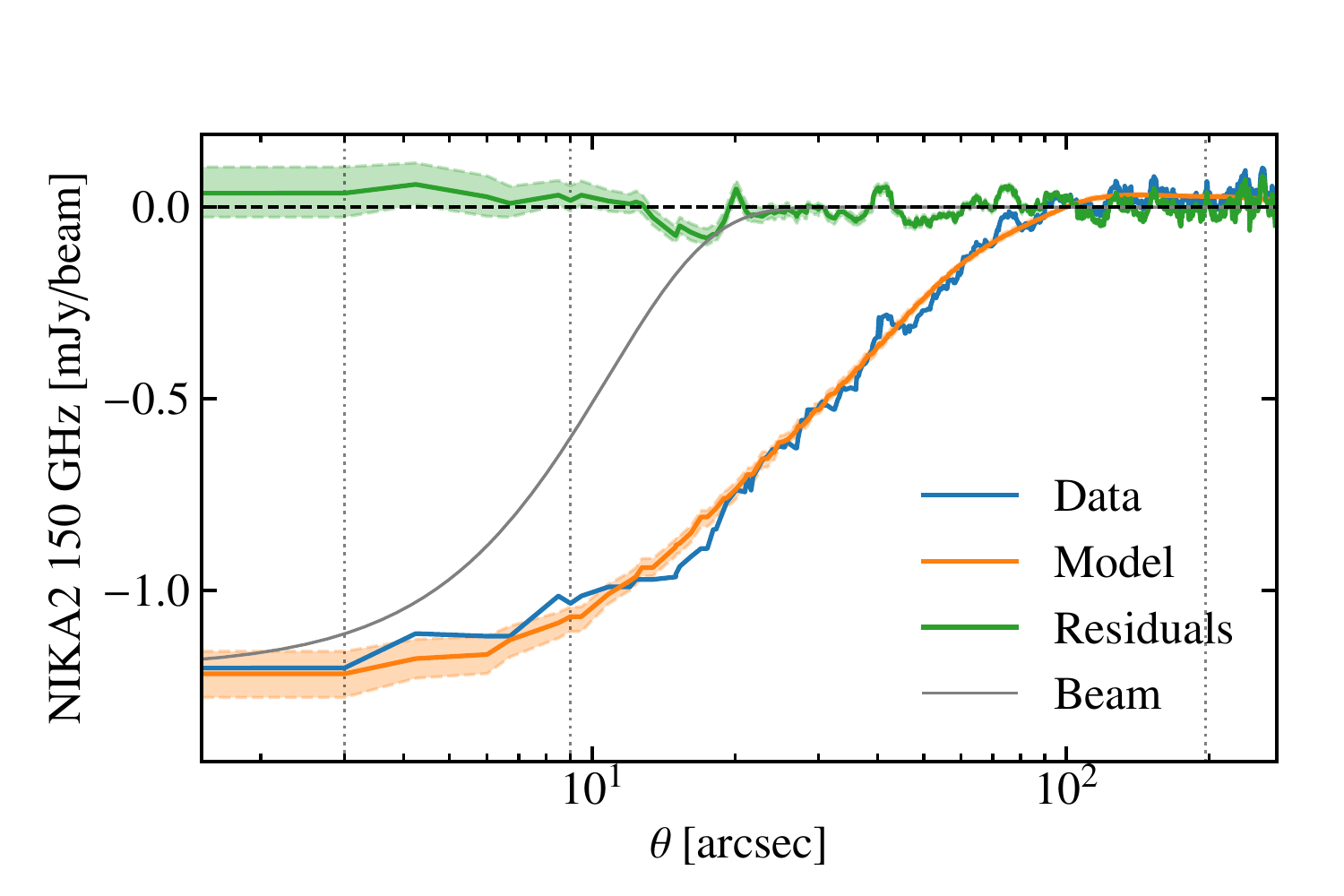}
    \includegraphics[height=4cm, trim={0 0 0 1cm}, clip]{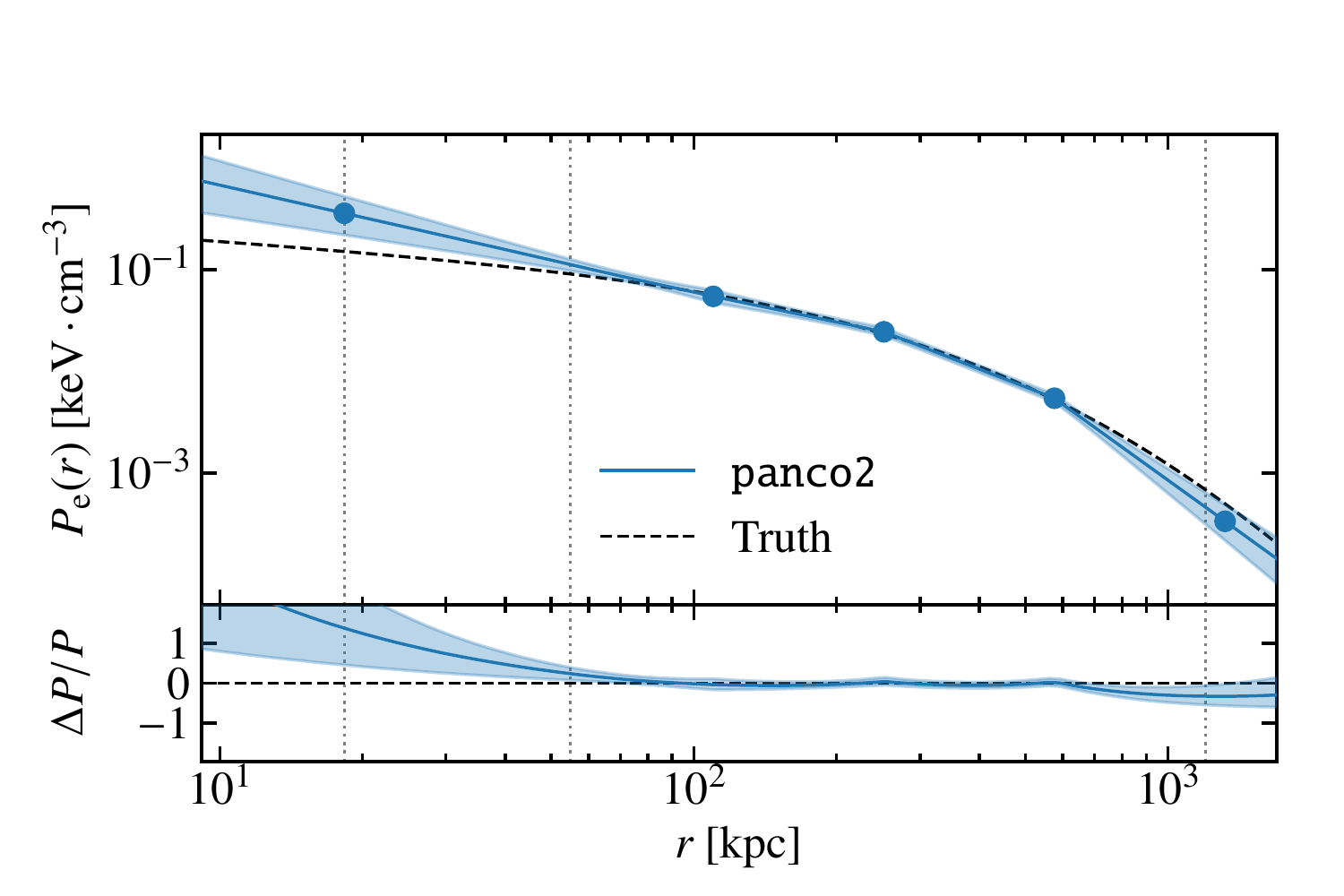}
    \caption{
        Validation results for the C2 cluster seen by NIKA2 with point source contamination, masking the point sources in the fit (see \S\ref{sec:simu:ps}).
        Legend is identical to Fig.~\ref{fig:valid2:tf2d}.
        Grey circles in the maps correspond to the pixels masked out during the analysis, located at the position of the injected point source contamination.
        A discrepancy in core pressure is noticeable between the true pressure profile of the cluster and the reconstruction by \panco, which can be attributed to a low amount of information in this region due to a lot of the pixels being masked and the projected radii being well inside the NIKA2 beam.
    }
    \label{fig:valid2:psmask}
\end{figure*}

\subsection{Integrated SZ signal constraint} \label{sec:simu:Ysz}

Finally, we run the fit of the mock NIKA2 map of the C2 cluster discussed in \S\ref{sec:simu} with an added constraint on its integrated tSZ signal, as described in \S\ref{sec:algo:likelihood}.
We use the $Y-M$ scaling relation of \citet{arnaud_universal_2010} (eq. 20) to compute an estimated value of $Y_{500} = 7.55 \times 10^{-5} \; {\rm Mpc^2}$ (within a radius $R_{500} = 1071.5 \; {\rm kpc}$).
We consider an uncertainty of 10\%.
The additional constraint is added to the log-likelihood using eq.~(\ref{eq:algo:likelihood_ysz}).

\begin{figure}[t]
    \centering
    \includegraphics[width=0.95\linewidth, trim={0 0 0cm 1cm}, clip]{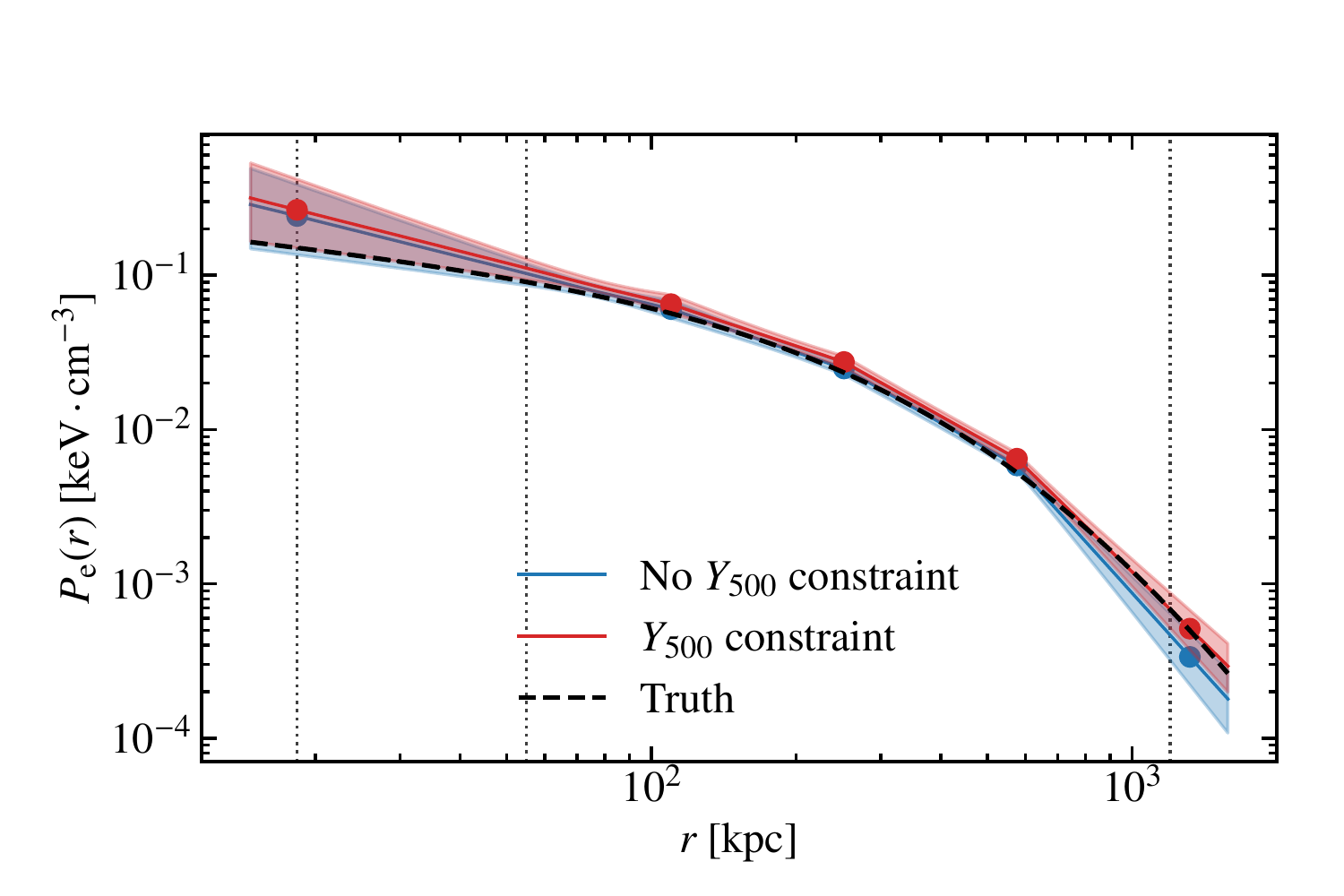}
    \caption{
        Pressure profile reconstructed by \panco\ from the NIKA2 mock map of the C2 cluster.
        In blue, the pressure profile reconstructed without a constraint on the integrated Compton parameter (identical to that shown in Fig.~\ref{fig:valid:profiles}).
        In red, the pressure profile reconstructed using a constraint on $Y_{500}$.
        Error envelopes the region between the 16th and 84th percentiles of the posterior distribution.
        Dotted vertical lines, from left to right, show the size of a map pixel, the beam HWHM, and the half map size.
        The dashed black line shows the true pressure profile used to generate the input map.
    }
    \label{fig:valid2:Y500}
\end{figure}

Figure \ref{fig:valid2:Y500} shows the reconstructed pressure profile, compared to the one previously extracted from the same map without considering the constraint on $Y_{500}$.
As expected, the difference between the profiles is minor in the inner region of the profile, in which the tSZ signal map dominates the constraint.
However, a difference arises at larger radii, where the constraining power of the map drops.
We can see that the $Y_{500}$ constraint has an impact on the reconstructed value of the last pressure bin, improving the agreement between the reconstructed profile and the truth in the outskirts of the cluster.
The reduced chi-squared value is given in table~\ref{tab:simu:chi2}, again showing goodness of fit.

\section{Discussion} \label{sec:discussion}

\subsection{Possible future improvements}

This release of \panco\ allows a user to perform the fit of a radially binned pressure profile model on a tSZ map.
Several additional features can be thought of to offer more flexibility in this kind of analysis.
Here, we offer a non-exhaustive list of possible such extensions.
We strongly emphasize that this list should \textit{not} be interpreted as a list of features currently in development by the authors, but instead as a collection of possible avenues to explore.

\paragraph{Joint analysis of several tSZ maps} 
We have shown in \S\ref{sec:simu} that \panco\ could recover pressure profiles from tSZ observations with different kinds of instruments -- namely, we have tested \textit{Planck}-, SPT-, and NIKA2-like mappings.
But our analysis treated all of these datasets separately.
For clusters with available data from several instruments, there may be interesting information to be extracted from the joint analysis of several maps \citep[see \eg][]{ruppin_first_2018, romero_multi-instrument_2018}.

\paragraph{Different profile parametrizations} 
As discussed in \S\ref{sec:algo:press}, many studies of pressure profiles from tSZ (or X-rays) observations use the gNFW profile.
Even though our radially-binned parametrization offers more flexibility and circumvents the known problems of the usual gNFW parametrization, one may be interested in using other functional forms of pressure profile.
For example, the rewriting of the gNFW model from \citet{battaglia_cluster_2012-1} \citep[see also \eg][]{sayers_evolution_2022} offers a way to constrain the same functional form, but minimizing the correlations in the parameter space.
Even though it is still possible to use a radially-binned profile and to fit any other functional form on the pressure bins \textit{a posteriori} \citep[see \eg][]{munoz-echeverria_multi-probe_2022}, one may want to directly fit said parametrization to the data, which is currently not supported.

\paragraph{Non-spherical pressure models} 
Galaxy clusters are well known to not be perfectly spherical structures.
Consequently, fitting a spherically-symmetric pressure distribution misses some information potentially contained in tSZ maps.
For example, clusters can be aspherical because they are an ongoing merger of two substructures.
Observations of such systems can give precious insights on the dynamics of cluster mergers, and therefore on the physics of large-scale structure formation.
Additionally, non-merging clusters may be aspherical because of their connection to the cosmic web, which can provide interesting insights on cosmic filaments.
Therefore, the possibility to fit pressure distributions beyond spherical symmetry is interesting.
Depending on the morphology of the system, this may be done by fitting for an ellipsoidal pressure distribution \citep[\eg][]{sarazin_deep_2016}, or the sum of the contributions of two (possibly spherical) halos \citep[\eg][]{artis_psz2_2022}.

\paragraph{Joint tSZ--X-ray analysis} 
The \textit{bremsstrahlung} of hot electrons makes the ICM emit in the X-ray domain \citep[see \eg][for reviews]{bohringer_x-ray_2010, bohringer_x-ray_2013}.
This radiation carries information on the ICM that is very complementary to that offered by tSZ observations.
Namely, similarly to how tSZ signal is linked to the electron pressure in the ICM -- as seen in eq.~(\ref{eq:algo_ysz}) -- X-ray emission is linked to its (squared) electron density.
Moreover, for sufficiently deep observations, X-ray data can also be used to study the spectral distribution of the detected photons, enabling measurements of the ICM electron temperature.
The combination of X-ray and tSZ observations of clusters therefore offers a way to finely characterize ICM thermodynamics.
This complementarity can be exploited \textit{a posteriori}, \eg\ by combining results from \panco\ with density and temperature profiles obtained independently of X-ray data \citep[\eg][]{keruzore_exploiting_2020}, or through a joint fit of the thermodynamic properties \citep[\eg][]{castagna_joxsz_2020-1}.
The possibility of joint tSZ--X-ray fits could be added to \panco\ to take further advantage of this complementarity.

\paragraph{Other general additional considerations} 
Several adjustments could be made to \panco\ to enable its usage on even more general data.
For example, the data inputs could be adapted to also include curved-sky projections.
Support for arbitrary PSF filtering (\eg\ non-Gaussian or even asymmetrical beams) could also be of interest, to account for the impact of complex diffraction patterns on reconstructed signals.

\subsection{On the choice of radial binning}

As discussed in \S\ref{sec:algo:press}, the pressure model fitted on the data requires the user to define a set of radii used as nodes for the power-law interpolation of the profile.
The choice of these radii is far from straightforward, as it depends on the angular scales present in the data to be fitted, as well as on the scientific goals of the analysis.
This is the main downside of the radially binned model used in \panco, as the radii chosen may have a significant impact on the analysis results.
The model dependence of products derived from radially-binned pressure profiles has been discussed by \citet{munoz-echeverria_multi-probe_2022}, who show that hydrostatic mass estimates, linked to the first derivative of the pressure profile, are particularly affected.

In order to try to circumvent this shortcoming, we strongly advise \panco\ users to be cautious about their choice of binning, and, when possible, to try different binnings.
We have implemented functions in \panco\ that allow users to easily simulate a mock tSZ map mimicking the data to be fitted.
We encourage users to use this functionality to create such datasets and try out their choices of radial binning on these.
Similarly to the validation presented in \S\ref{sec:simu:results}, the comparison of the pressure profile reconstructed with the one used to generate the map will provide insights on how well adapted a choice of radial binning is to a dataset.
How to perform such an analysis is presented in the accompanying technical documentation.
Investigations towards the systematic impact of radial binning on recovered pressure profiles will be the subject of a future study.

\section{Conclusion} \label{sec:conclusion}

This paper presents the release of \panco, a software allowing its users to perform pressure profile extraction from a tSZ map.
We have highlighted the main features of the software and its algorithm, based on forward modeling MCMC of the tSZ signal with a radially-binned pressure profile.
We have presented the variety of observational systematics that can be incorporated in the modeling in order to account for the known features of millimeter-wave observations.
We have validated the software and its different functionalities on realistic simulated maps of different clusters.

Our main conclusions are as follow.

\begin{itemize}[leftmargin=*]
    \item Using realistic simulations, we have shown that \panco\ could be used to recover accurate pressure profiles from different projections of tSZ mappings.
        In particular, \S\ref{sec:simu} showed validation on three clusters, covering different parts of the mass-redshift plane, mapped to mimic the \textit{Planck} $y-$map of \citet{planck_collaboration_planck_2016-2}, the SPT $y-$map of \sptymap, and realistic NIKA2 150 GHz observations, effectively covering a range of angular resolutions from $18''$ to $10'$.
    \item We have demonstrated the possibility to include correlated noise (\S\ref{sec:simu:corr_noise}) and anisotropic filtering (\S\ref{sec:simu:2dtf}) in the analysis, and still recover accurate pressure profiles.
    \item We have shown two different ways to account for point source contamination in \panco, by masking the sources or marginalizing over their contamination (\S\ref{sec:simu:ps}).
\end{itemize}

The combination of all these features and their validation makes \panco\ a flexible and robust software, capable of retrieving accurate pressure profile estimations from a large variety of tSZ maps.

\subsection{Products released}

The entirety of the \panco\ software is made available as a github repository: \\
\url{https://github.com/fkeruzore/panco2}. \\
The software version described in this work is \texttt{v1.0}.
That same repository also includes the data generated for the validation described in \S\ref{sec:simu} and \S\ref{sec:extra_simu}, along with the products of said validation (\ie\ the Markov chains and the running sequences used for each dataset).

In addition, \panco\ is also accompanied by an online technical documentation at: \\
\url{https://panco2.readthedocs.io}. \\
It includes detailed explanations of the technical aspects of the code, as well as a description of the inputs and outputs of \panco's different functions.
The documentation also gives examples of analyses that can be performed with \panco\ beyond the ones presented in the validation.
\vspace*{10pt}

\section*{Acknowledgements}

\small
We thank Lindsey Bleem for useful discussions and advice, in particular regarding SPT-like data.
The development of \panco\ comes after many years of work on pressure profile fitting codes within the NIKA and NIKA2 collaborations, for which we thank the NIKA2 collaboration, and acknowledge major contributions by Rémi Adam, Charles Romero, and Matthew McWilliam.
Argonne National Laboratory's work was supported by the U.S. Department of Energy, Office of Science, Office of High Energy Physics, under contract DE-AC02-06CH11357.
This work is supported by the French National Research Agency in the framework of the ``Investissements d'avenir'' program (ANR-15-IDEX-02).
\normalsize


\bibliographystyle{aasjournal}
\bibliography{panco2}


\appendix

\section{Posterior plots for validation fits} \label{sec:ap:corner}

\begin{figure*}[t]
    \centering
    \includegraphics[width=0.8\linewidth, trim={1cm 1cm 1cm 1cm}, clip]{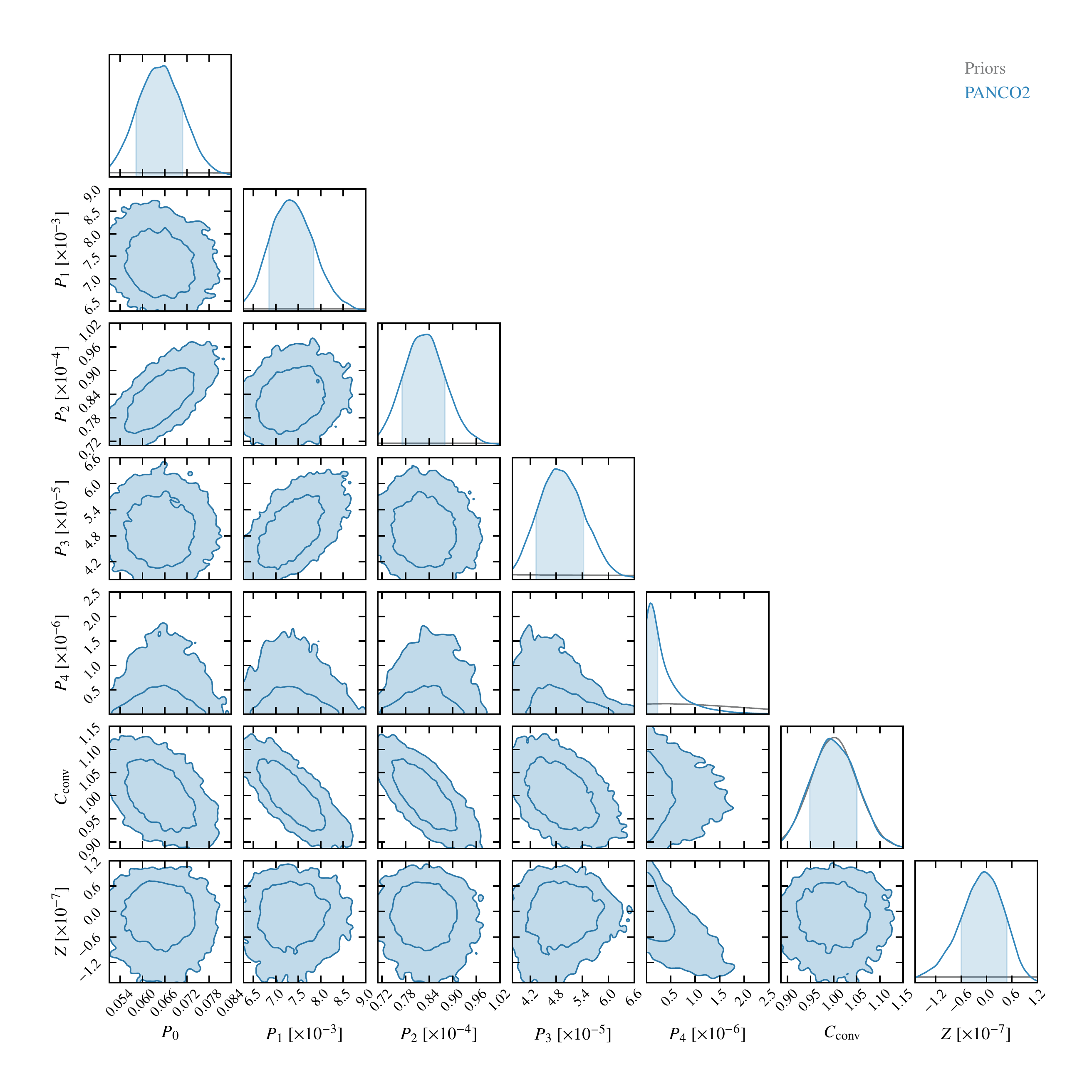}
    \caption{
        Posterior sampled during the fit of the C1 \textit{Planck} validation map.
        The diagonal shows the marginalized posterior on individual parameters, while the extra-diagonal elements show 2D projections.
        The shaded area in the 1D marginalized distributions show the $1\sigma$ constraints on each parameter, while the 2D contours are $1\sigma$ and $3\sigma$.
        For 1D parameter distributions, we also show the prior distribution in grey (see text).
    }
    \label{fig:app:corner1}
\end{figure*}

\begin{figure*}[t]
    \centering
    \includegraphics[width=0.8\linewidth, trim={1cm 1cm 1cm 1cm}, clip]{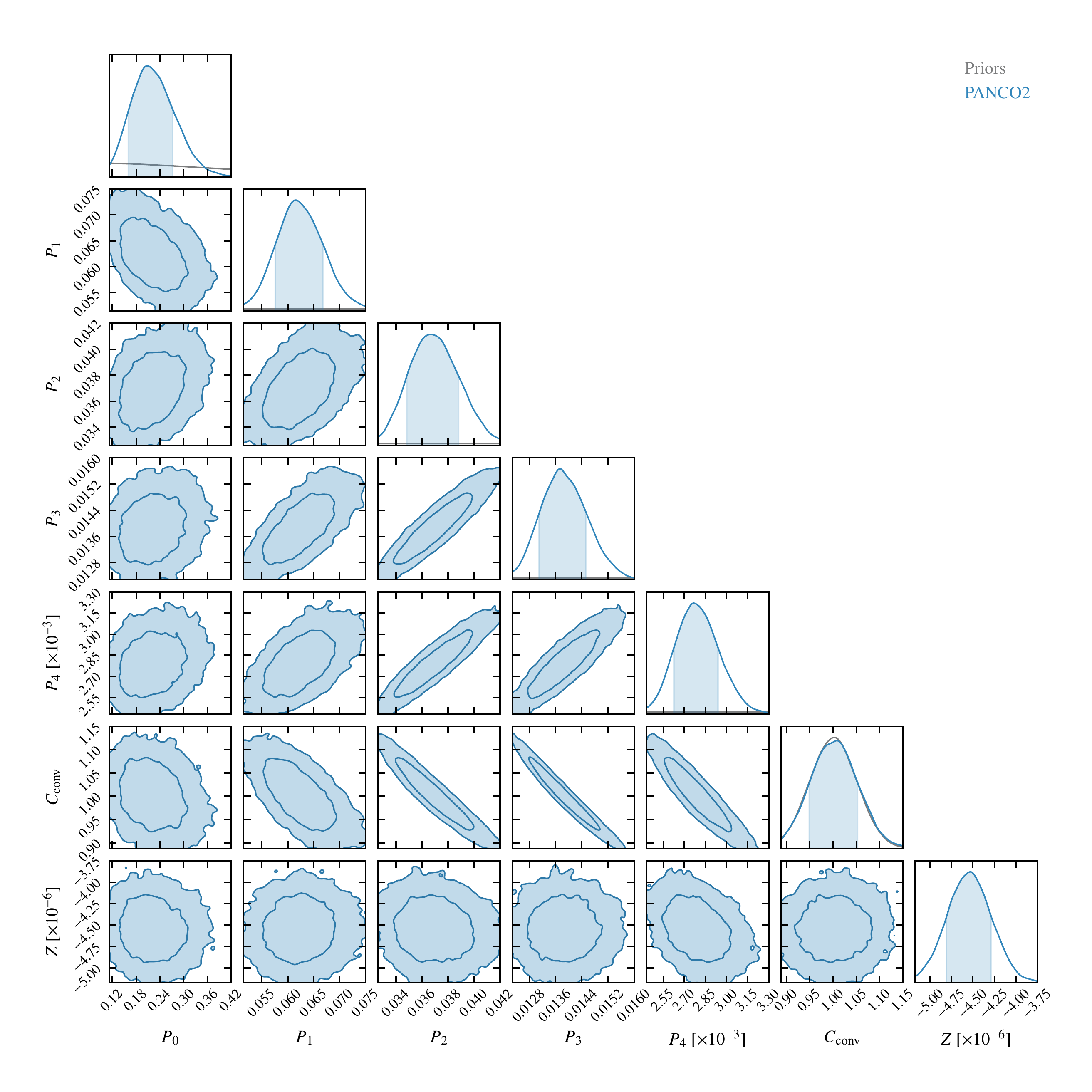}
    \caption{
        Posterior sampled during the fit of the C1 SPT validation map.
        Legend is identical to Figure~\ref{fig:app:corner1}
    }
    \label{fig:app:corner2}
\end{figure*}

\begin{figure*}[t]
    \centering
    \includegraphics[width=0.8\linewidth, trim={1cm 1cm 1cm 1cm}, clip]{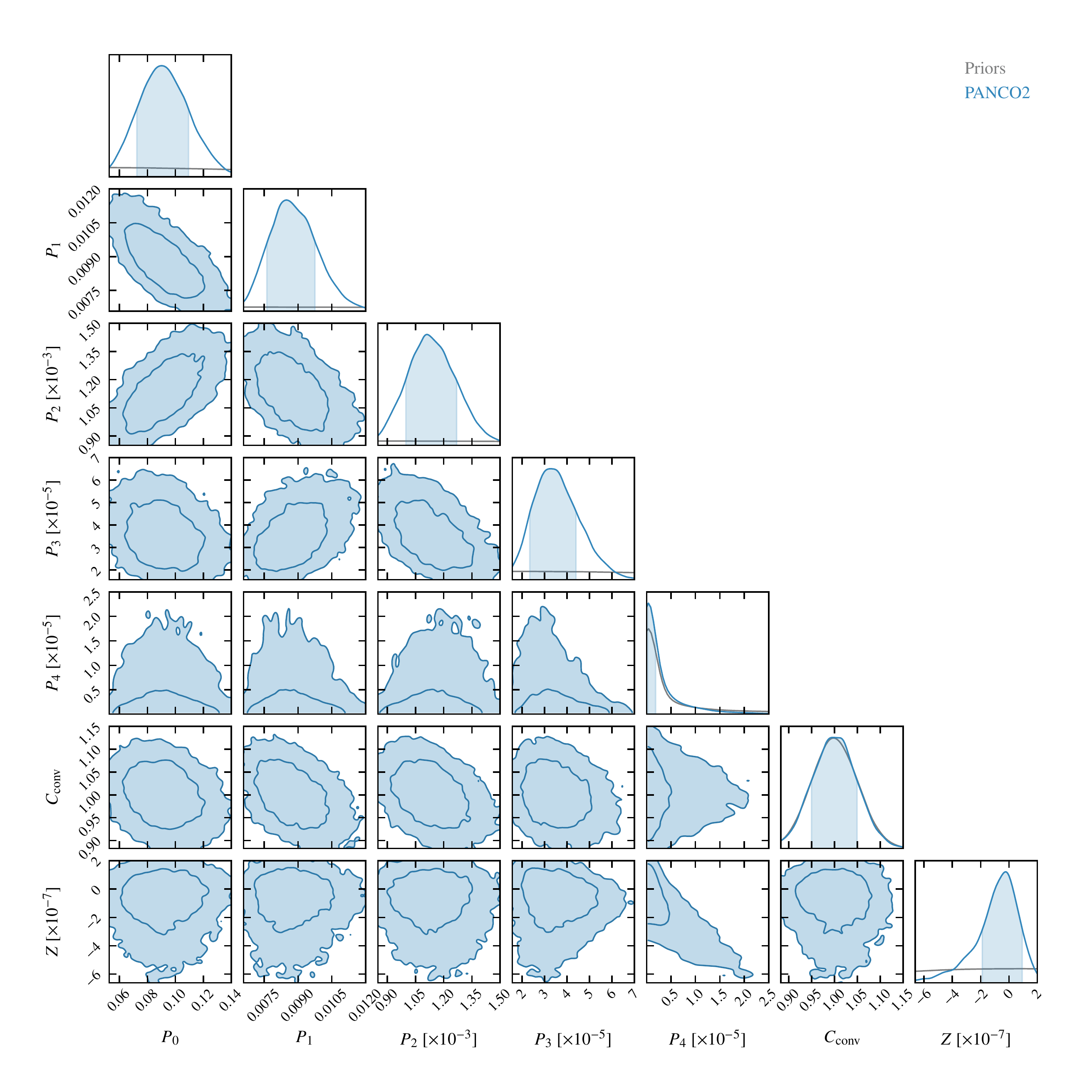}
    \caption{
        Posterior sampled during the fit of the C2 SPT validation map.
        Legend is identical to Figure~\ref{fig:app:corner1}
    }
    \label{fig:app:corner3}
\end{figure*}

\begin{figure*}[t]
    \centering
    \includegraphics[width=0.8\linewidth, trim={1cm 1cm 1cm 1cm}, clip]{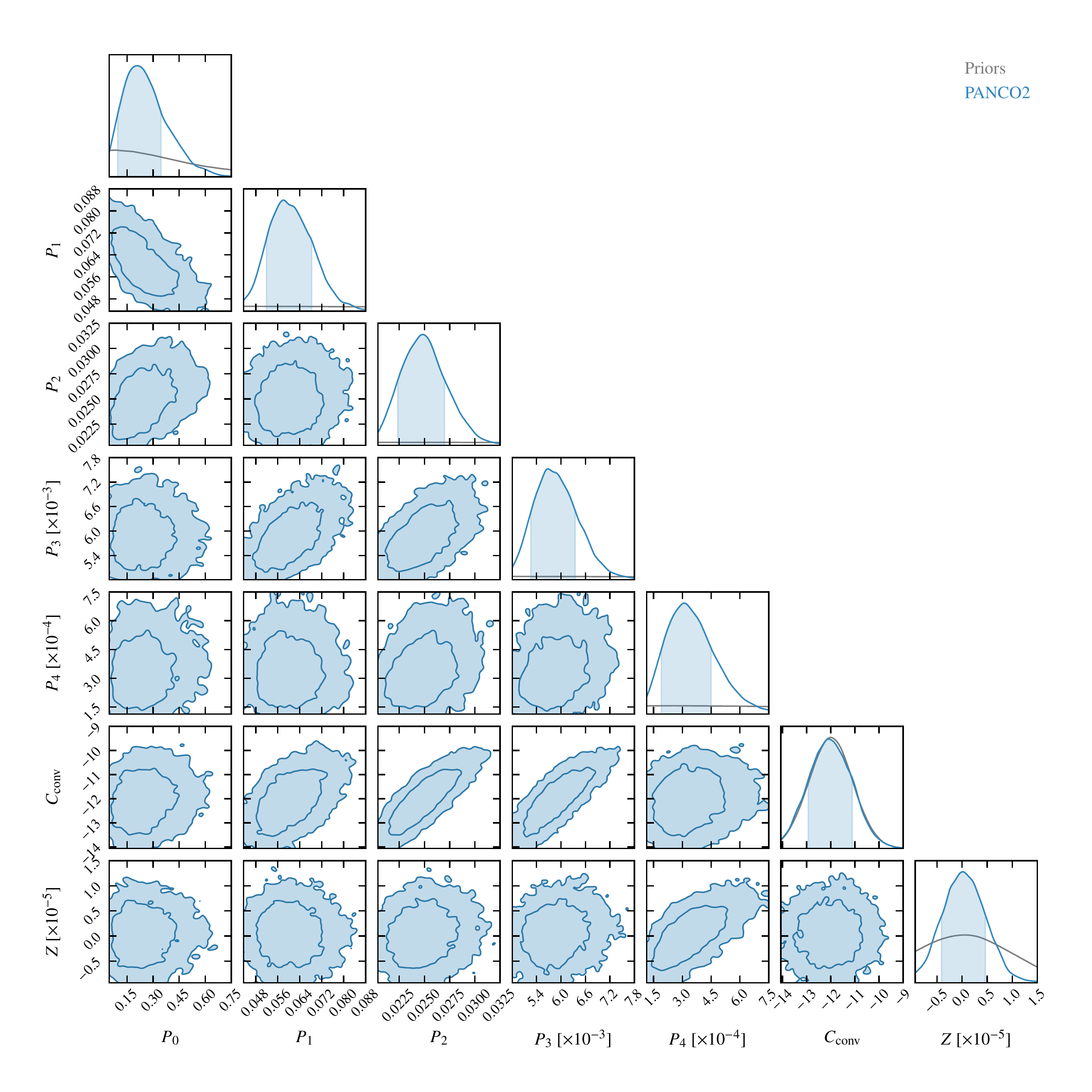}
    \caption{
        Posterior sampled during the fit of the C2 NIKA2 validation map.
        Legend is identical to Figure~\ref{fig:app:corner1}
    }
    \label{fig:app:corner4}
\end{figure*}

\begin{figure*}[t]
    \centering
    \includegraphics[width=0.8\linewidth, trim={1cm 1cm 1cm 1cm}, clip]{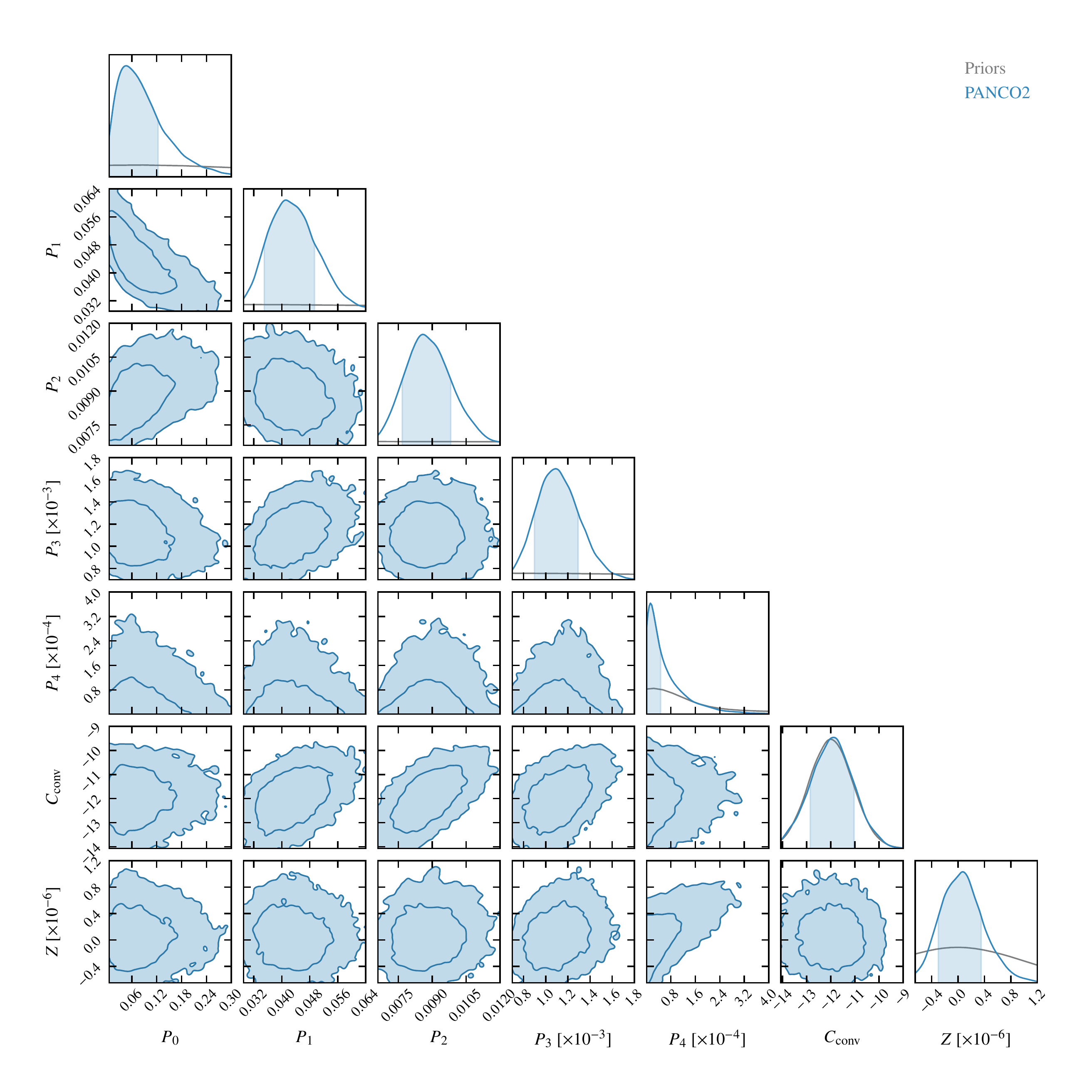}
    \caption{
        Posterior sampled during the fit of the C3 NIKA2 validation map.
        Legend is identical to Figure~\ref{fig:app:corner1}
    }
    \label{fig:app:corner5}
\end{figure*}

In order to further detail the results of the fits performed in the validation, we show in Figures~\ref{fig:app:corner1} to \ref{fig:app:corner5} the posterior sampled by our MCMC procedure in the parameter space.
We show them specifically for the five validation fits shown in \S\ref{sec:simu}.
For each figure, we show the results in the entire parameter space (described in \S\ref{sec:algo:likelihood}).
The posterior plots can be used to judge the correlation between different analysis parameters.
We note the general pattern of successive bins being anti-correlated, which is well known for this type of deprojections \citep[see \eg][]{sarazin_deep_2016, romero_multi-instrument_2018}.
We also notice an (anti-)~correlation between the conversion coefficient $C_{\rm conv}$ and the pressure bins, arising from the degeneracy between this parameter and the amplitude of the pressure profile.
Finally, we notice an (anti-)~correlation between the pressure in the last bin and the zero level of the map, due to the fact that changes in pressure in the outskirts of the cluster tend to manifest as a change in the signal at large angular scales, \ie\ similar to a constant offset.

We also show the marginalized prior distributions to illustrate the weight of priors in the analysis.
As seen in all the plots, the shape of the marginalized posterior is vastly different to that of the priors, showing that our fit results are not prior-dominated.
The only exception is the conversion coefficient $C_{\rm conv}$, for which the prior and posterior are identical.
This similarity simply indicates that the data do not constrain the parameter, which was expected, since it is a nuisance parameter, included in the fit to propagate uncertainties on the calibration of the map to the pressure profile estimate.
\vspace{15pt} 


\end{document}